\documentclass{aa}  

\usepackage[hidelinks,colorlinks=true,linkcolor=blue,citecolor=blue]{hyperref}
\usepackage{amsmath}    % Advanced maths commands
\usepackage{amssymb}    % Extra maths symbols
\usepackage{multirow}
\usepackage{booktabs}
\usepackage{inputenc}
\usepackage{siunitx}
\usepackage{hyperref}
\usepackage{graphicx}
\usepackage{txfonts}

\usepackage[dvipsnames]{xcolor}
\definecolor{BAROSI}{rgb}{0, 70.2, 70.2}

\newcommand{\hi}{H{\sc i}}

\begin{document}

   \title{The BINGO project IV:}
    \subtitle{Simulations for mission performance assessment and preliminary component separation steps}

\author{
Vincenzo Liccardo\thanks{E-mail: vic2000@hotmail.it}\inst{1},
Eduardo J. de Mericia\inst{1},
Carlos A. Wuensche\inst{1},
Elcio Abdalla\inst{2},
Filipe B. Abdalla\inst{1,2,3,4},
Luciano Barosi\inst{5},
Francisco A. Brito\inst{5,15},
Amilcar Queiroz\inst{5},
Thyrso Villela\inst{1,6},
Michael W. Peel\inst{13,14},
Bin Wang\inst{7,11},
Andre A. Costa\inst{7},
Elisa G. M. Ferreira\inst{2,8},
Karin S. F. Fornazier\inst{2},
Camila P. Novaes\inst{1},
Larissa Santos\inst{7,11},
Marcelo V. dos Santos\inst{5},
Mathieu Remazeilles\inst{9},
Jiajun Zhang\inst{12},
Clive Dickinson\inst{9},
Stuart Harper\inst{9},
Ricardo G. Landim\inst{10},
Alessandro Marins\inst{2},
Frederico Vieira\inst{1}
}

%1
\institute{
Divis\~ao de Astrof\'isica, Instituto Nacional de Pesquisas Espaciais - INPE, Av. dos Astronautas 1758, 12227-010 - S\~ao Jos\'e dos Campos, SP, Brazil
    \and %2
Instituto de F\'isica, Universidade de S\~ao Paulo,  05315-970, S\~ao Paulo, Brazil
 \and %3
University College London, Gower Street, London,WC1E 6BT, UK 
    \and %4
Department of Physics and Electronics, Rhodes University, PO Box 94, Grahamstown, 6140, South Africa
    \and %5
Unidade Acad\^emica de F\'{i}sica, Universidade Federal de Campina Grande, R. Apr\'{i}gio Veloso, 58429-900 - Bodocong\'o, Campina Grande - PB, Brazil 
 \and %6
Instituto de F\'{i}sica, Universidade de Bras\'{i}lia, Campus Universit\'ario Darcy Ribeiro, 70910-900, Bras\'{i}lia, DF, Brazil 
    \and %7
Center for Gravitation and Cosmology, College of Physical Science and Technology, Yangzhou University, Yangzhou 225009, China
\and %8
Max-Planck-Institut f{\"u}r Astrophysik, Karl-Schwarzschild Str. 1, 85741 Garching, Germany
     \and %9
Jodrell Bank Centre for Astrophysics, Department of Physics and Astronomy, The University of Manchester, Oxford Road, Manchester, M13 9PL, U.K. 
    \and %10
Technische Universit\"at M\"unchen, Physik-Department T70, James-Franck-Stra\text{$\beta$}e 1, 85748 Garching, Germany
    \and %11
School of Aeronautics and Astronautics, Shanghai Jiao Tong University, 200240, China
    \and %12
Center for Theoretical Physics of the Universe, Institute for Basic Science (IBS), Daejeon 34126, Korea
        \and %13
Instituto de Astrof\'{i}sica de Canarias, 38200, La Laguna, Tenerife, Canary Islands, Spain
    \and %14
Departamento de Astrof\'{i}sica, Universidad de La Laguna (ULL), 38206, La Laguna, Tenerife, Spain 
    \and %15
Departamento de Física, Universidade Federal da Paraíba, Caixa Postal 5008, 58051-970 João Pessoa, Paraíba, Brazil
}

% \abstract{}{}{}{}{} 
% 5 {} token are mandatory
 
  \abstract
  % context heading (optional)
  % {} leave it empty if necessary  
  % aims heading (mandatory)
   {}
  % methods heading (mandatory)
   {The large-scale distribution of neutral hydrogen (\textsc{Hi}) in the Universe is luminous through its 21 cm emission. The goal of the \textbf{B}aryon Acoustic Oscillations from \textbf{I}ntegrated \textbf{N}eutral \textbf{G}as \textbf{O}bservations -- \textbf{BINGO} -- radio telescope is to detect baryon acoustic oscillations (BAOs) at radio frequencies through 21 cm intensity mapping (IM). The telescope will span the redshift range 0.127 $< z <$ 0.449 with an instantaneous field-of-view of $14.75^{\circ} \times 6.0^{\circ}$.}  
  % results heading (mandatory)
   {In this work we investigate different constructive and operational scenarios of the instrument by generating sky maps as they would be produced by the instrument. In doing this we use a set of end-to-end IM mission simulations. The maps will additionally be used to evaluate the efficiency of a component separation method ({\tt GNILC}). }
  % conclusions heading (optional), leave it empty if necessary 
   {We have simulated the kind of data that would be produced in a single-dish IM experiment such as BINGO. According to the results obtained, we have optimized the focal plane design of the telescope. In addition, the application of the {\tt GNILC} method on simulated data shows that it is feasible to extract the cosmological signal across a wide range of multipoles and redshifts. The results are comparable with the standard principal component analysis method.}
   {}


    \date{Received XX, XX, 2020; accepted XX, XX, 2020} 

    \titlerunning{The BINGO project IV: Mission Simulations }
    \authorrunning{V. Liccardo et al.}

   \maketitle

%%%%%%%%%%%%%%%%% BODY OF PAPER %%%%%%%%%%%%%%%%%%

\section{Introduction}

Constraints on the nature and properties of dark energy can be obtained from an accurate description of the Universe's expansion history. Such a description can be provided by standard rulers: objects or properties of known size for which we can easily retrieve the distance-redshift relation \citep{Weinberg:2013}. Baryon acoustic oscillations (BAOs), frozen relics of the epoch when matter and radiation were coupled together, are promising standard ruler candidates and could help us understand more about the nature of dark energy  \citep[see][]{Albrecht06}. The current scenario is that, for much of cosmic history, matter dominated over dark energy and the expansion indeed slowed, enabling galaxies and large-scale structures (LSSs) to form. A billion years ago, matter became sufficiently dilute due to expansion, dark energy became the dominant component of the Universe, and the expansion accelerated.

To date, BAOs have only been detected by performing large galaxy redshift surveys in the optical waveband \citep{Eisenstein05}. However, given the implications of these measurements, it is important that they be confirmed in other wavebands and measured over a wide range of redshifts. The radio band provides a unique and complementary observational window for the understanding of dark energy via the redshifted 21 cm  neutral hydrogen emission line from distant galaxies. The \textbf{B}aryon Acoustic Oscillations from \textbf{I}ntegrated \textbf{N}eutral \textbf{G}as \textbf{O}bservations (BINGO) telescope is a proposed new instrument designed specifically to observe such a signal and to provide a new insight into the Universe \citep{Battye13,2020_project}. 

The telescope design consists of two dishes in a compact configuration with no moving parts and will operate in the frequency range from 0.98\,GHz to 1.26\,GHz (corresponding to a co-moving distance of 380--1280\,Mpc/$h$ assuming a $\Lambda$CDM cosmology; \citealt{Planck2018a}). It will map the cosmic web in three dimensions without detecting individual galaxies, a technique called intensity mapping  \citep[IM;][]{Peterson06}. Instead of cataloging many individual galaxies, one can study the LSS directly by detecting the aggregate emission from many galaxies that occupy large $\approx$ 100 Mpc$^{3}$ voxels. The unresolved 21 cm  signal from the galaxies is therefore similar to a low-resolution galaxy survey and can be used as a low-z cosmological probe via BAO measurements. The full-width at half-maximum of the BINGO beam is 40\,arcmin, allowing structures of angular sizes corresponding to a linear scale of around 150\,Mpc to be resolved (BAOs manifest themselves as a small but detectable excess of galaxies with separations of \mbox{$\approx$ 150\,Mpc} in the chosen redshift range). 

Large-scale \textsc{Hi} fluctuations above redshift $z$ = 0.1 have been unambiguously detected only in cross-correlation with existing surveys of optically selected galaxies \citep{Lah09, Chang10, Masui13}. Cross-correlation between the cosmological 21 cm signal and optical surveys provides potentially useful information on the statistical properties of the \textsc{Hi} distribution \citep{Switzer15}.
%a potentially useful statistics. 
The cross-correlation has the advantage, in comparison to ``autocorrelation'' studies, that the measured statistics are less sensitive to contaminants such as foregrounds, systematics, and noise.
The detection of the redshifted 21 cm radiation will provide valuable information about the post-recombination history of the Universe, including the Dark Ages. Information can also be extracted about the formation of the first ionizing sources and the subsequent reionization of the intergalactic medium (IGM) due to these sources (for a comprehensive review, see \citealt{Furlanetto06} and \citealt{Pritchard12}). 

Broadband foreground emission poses the greatest challenge to 21 cm IM and needs to be characterized carefully before the technique becomes a sensitive probe of the post-recombination epoch ($z < 160$). The foregrounds are expected to be predominantly Galactic and approximately four orders of magnitude larger than the cosmological signal (at low frequencies, synchrotron emissions from our Galaxy and other radio galaxies are the dominant foregrounds; \citealt{Sazy15}). Therefore, one of the key observational challenges in detecting the cosmological 21 cm signal is modeling and removing Galactic foregrounds at low frequencies. In doing this, the BINGO pipeline adopts the component separation strategies that use the spatial structure of foregrounds to separate them from the cosmological signal. Indeed, the cosmological 21 cm signal is expected to have structure in frequency space, while the foregrounds are expected to mostly be spectrally smooth. 
Many of these techniques were first developed for cosmic microwave background (CMB) data analysis and are now being extended to IM, where there is the extra dimension of frequency \citep{Ansari12, Switzer13}.
There are still spectrally un-smooth emission sources, such as radio frequency interference (RFI) from terrestrial and non-terrestrial sources, that can dominate over Galactic and extragalactic foregrounds. Radio frequency interference can be minimized through software removal, by choosing radio-quiet locations, and through band selection.

This work aims to test and optimize the constructive and operational parameters of the telescope, as well as the data analysis process itself. A set of computational routines and procedures (pipeline) that simulate the BINGO operation has been implemented. Its input is composed of maps of different emission mechanisms, produced by theoretical models or observations, and inherent noise properties of the equipment and the environment. The number and arrangement of horns, optical design, and receiver characteristics are also input parameters of the radio telescope. The IM pipeline produces, as output, time-ordered data sets (TODs) and antenna temperature maps that simulate the signal picked up by the instrument during a given period of operation. Next, these output data are passed through a component separation process for the recovery of the  \textsc{Hi} component.

The paper proceeds as follows: In  Sect. \ref{sec:instr} we give an overview of the instrument. In Sect. \ref{sec:simulations} the pipeline is briefly introduced along with a complementary discussion of the input models, the latest configuration updates, and the component separation method ({\tt GNILC}). Then, in Sect. \ref{sec:strategy} we study the observational efficiency of different feed horn arrangements. This is followed by the simulations and the component separation results and analysis (Sect. \ref{sec:gnilc}). Finally, we present the conclusion, with several future prospects.

 Throughout this paper we use cosmological parameters from \cite{Planck2018a}. This is the fourth (IV) of a series of papers describing the BINGO project. The theoretical and instrumental projects are in papers I and II  \citep{2020_project,2020_instrument}, the optical design in paper III \citep{2020_optical_design}, the component separation and correlations in paper V \citep{2020_component_separation}, the simulations for a mock 21 cm catalog are described in paper VI \citep{2020_mock_simulations}, and the cosmological forecasts for BINGO in paper VII \citep{2020_forecast}.

%%%%%%%%%%%%%%%%%  
\section{The instrument}
\label{sec:instr}

The telescope will be built on a hill near Aguiar, Para\'{\i}ba (northeastern Brazil). Earlier concepts of the BINGO can be found in \cite{Battye13} and \cite{Wuensche:2018}. The primary dish will be a 40\,m diameter paraboloid and the secondary, a 34\,m-diameter hyperboloid. The particular mirror configuration chosen is the crossed Dragone, also known as Compact Range Antenna. Such a design has very low geometric aberrations, leading to an instantaneous field-of-view of $\approx$ 88 deg$^{2}$. It will provide excellent polarization performance and very low side-lobe levels required for \textsc{Hi} IM. %Figure \ref{fig:design} presents the current BINGO optical schematics. 
A detailed description of the project and the instrument is available in companion papers I and II \citep{2020_project,2020_instrument}.

%\begin{figure}
    %\centering
        %\includegraphics[width=\columnwidth]{figures/design.jpg}
    %\caption{Optical design of the BINGO telescope.}
    %\label{fig:design}
%\end{figure}

\begin{table}
\centering
\caption{Summary of the BINGO Phase 1 telescope parameters.} 
% \vspace{2pc}
% \resizebox{6cm}{!}{
\begin{tabular}{c||c} \toprule 
{\bf Description} & {\bf Value} \\ \midrule \midrule                               
Dish diameters (m) & 40 (primary) \\ %\midrule
 & 34 (secondary)  \\ \midrule
Resolution ($^{\circ}$) & $\approx$ 0.67 \\ \midrule
Focal length (m) & 63.2  \\ \midrule
Frequency range (MHz) & 980--1260 \\ \midrule 
Channel resolution (MHz) &  9.33    \\ \midrule
Z interval  & 0.127--0.449 \\ \midrule
Number of feeds $n_{f}$ & 28 \\ \midrule
Central focal plane array elevation ($^{\circ}$) & $\approx$ 82 \\ \midrule
Azimuth ($^{\circ}$) & 0 (North) \\ \midrule
%Telescope area (m$^{2}$) & 1602 \\ \midrule
Telescope effective area (m$^{2}$) & $\approx 1120$ \\ \midrule
Pixel solid angle (sr) - $\Omega_{pix}$ & 0.35 \\ \midrule
Field of view (deg$^{2}$) & 14.75 $\times$ 6.0 \\ \midrule
Survey area $\Omega_{sur}$ (deg$^{2}$) & $\approx$ 5324 \\ \midrule
System temperature $T_{\textsc{sys}}$ (K)  &  $\approx$ 70 \\   \bottomrule  \bottomrule
\end{tabular} 
% } 
\label{tab_parameters}
\end{table}

\begin{figure}[t]
    \centering
        \includegraphics[height=8.5cm]{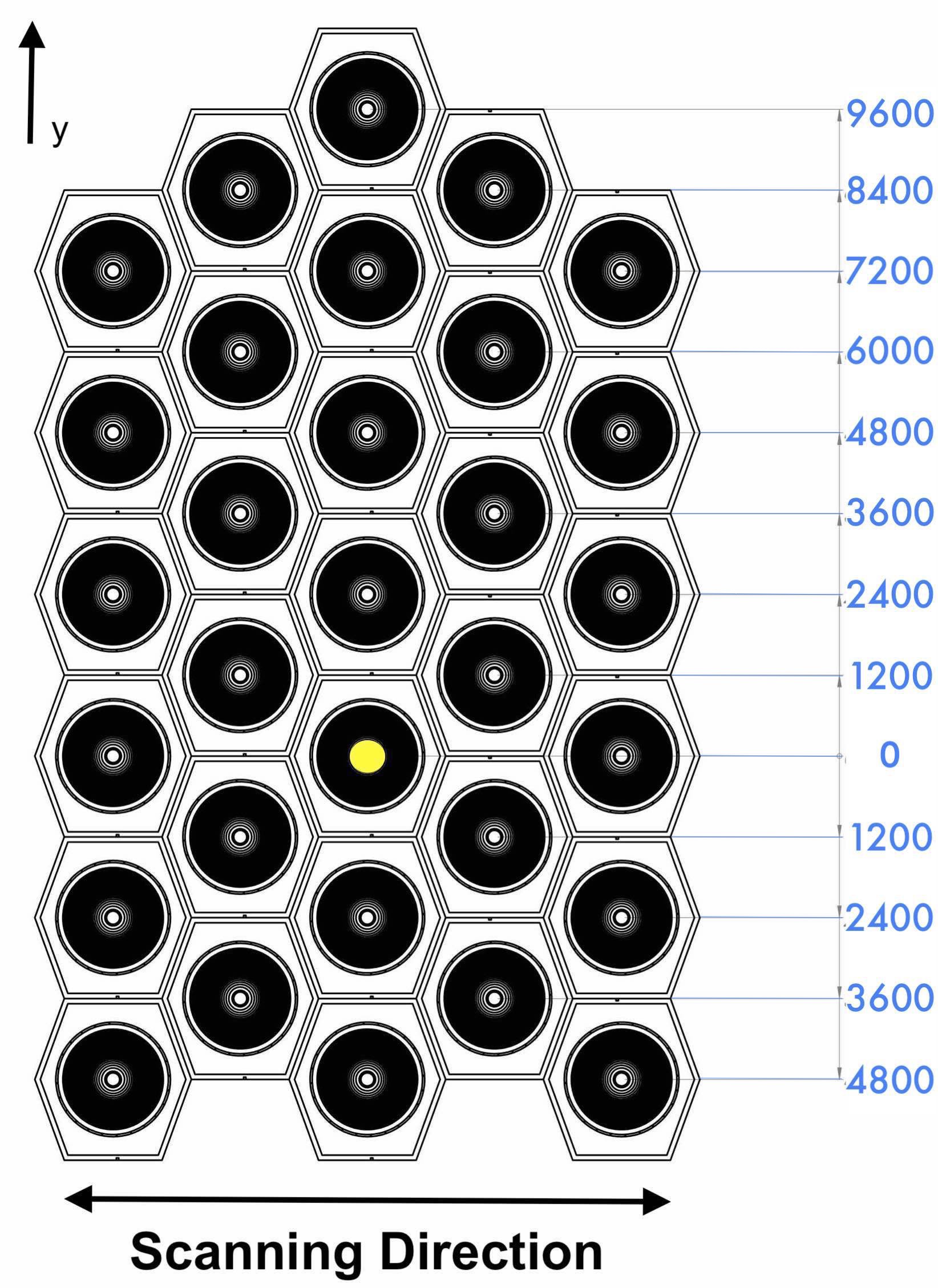}
    \caption{Hexagonal feed horns array ($\approx$ 9\,m $\times$ 17\,m). The numbers represent the positions of the horns along the $y$ axis, from the center of the field-of-view, given in mm.}
    \label{fig:arranjo_31}
\end{figure}

\begin{figure}[t]
    \centering
        \includegraphics[height=8.5cm]{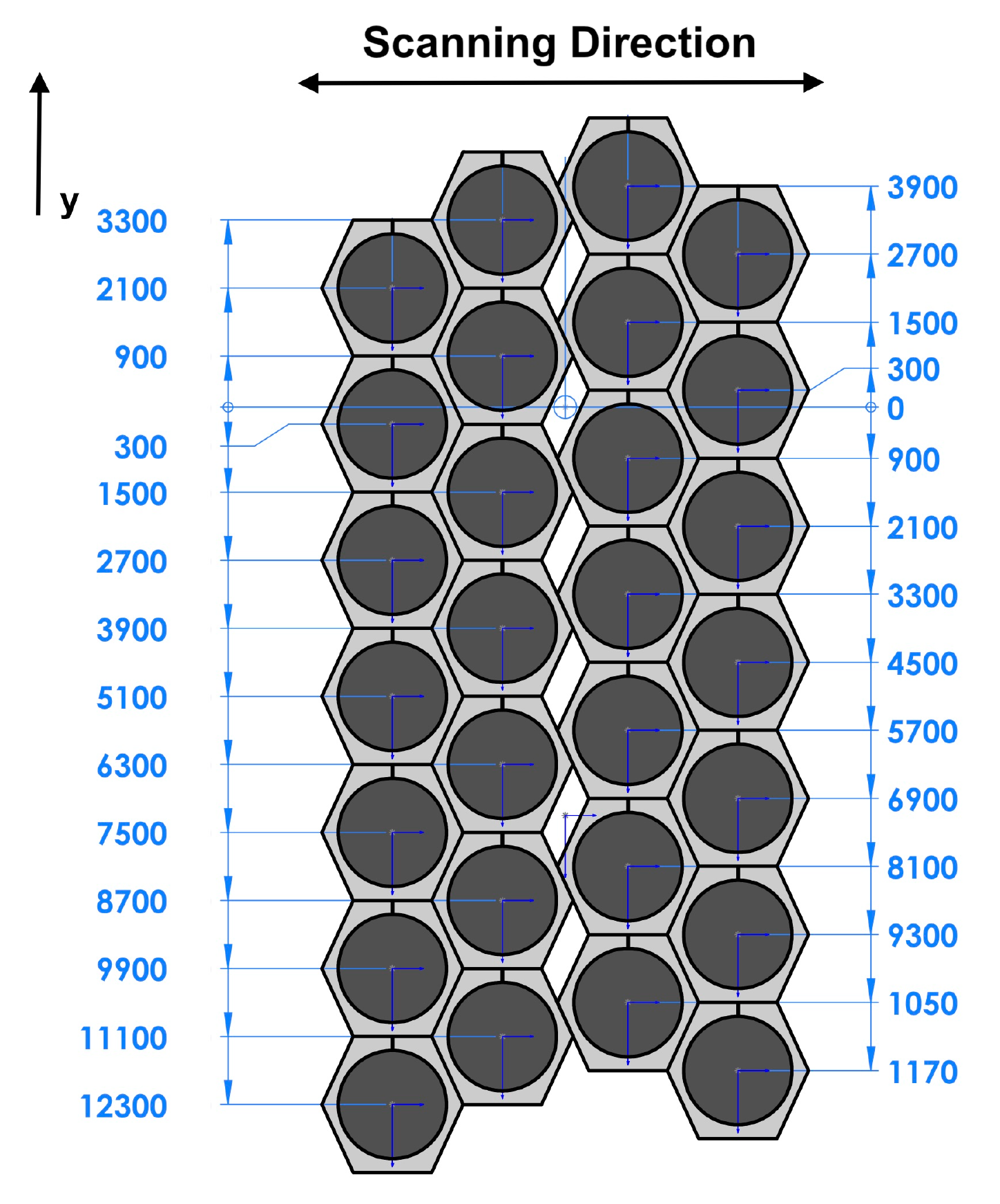}
    \caption{Double-rectangular feed horns array ($\approx$ 7.8\,m $\times$ 18.6\,m). The numbers represent the positions of the horns along the $y$ axis, from the center of the field-of-view, given in mm.}
    \label{fig:arranjo_28}
\end{figure}

In this paper we compare the two feed horns arrays considered in the current optical design \citep[see][]{2020_optical_design}: the first one (hexagonal) is composed of 31 units (Fig. \ref{fig:arranjo_31}), 
while the second one (double-rectangular) is made up of 28 units 
(Fig. \ref{fig:arranjo_28}). 
One of the purposes of this work is to investigate, through simulations, which  arrangement best meets the scientific goals defined for BINGO.

Each horn will be secured in the focal plane by a hexagonal casing, which works both as a transportation box and assembly cell. It will encapsulate the horn, transitions, polarizer, magic tee, and the receiver box \citep{2020_instrument}. With this hexagon concept, it is likely that no additional external structure will be needed to position the horn array structure. The hexagonal casing will be 2400 mm tall and will allow moving the horn in elevation $y$ and azimuth $x$ directions, as well as longitudinally $z$ (along the horn optical axis). By means of a pivot attached to the hexagonal case, where the horn is mounted, it will be possible to do a fine positioning of the horns. The aim is to reproduce the desired curvature of the focal plane by translating and tilting the horns across the array. This configuration will improve the sky coverage and final science products. In the future, the structure has the capacity to increase the number of horns to 56 for the double-rectangular array (by adding two extra columns on both sides) and 49 for the hexagonal one (by adding two columns on both sides with five and four horns, respectively, in such a way as to form a hexagon). This will double the redundancy (for the double-rectangular configuration) in the area covered and will increase the sensitivity by $\sqrt{2}$ per year.

%%%%%%%%%%%%%%%%% NEW SECTION %%%%%%%%%%%%%%%%%%

\section{Simulations and data processing}
\label{sec:simulations}

In this section we briefly describe the data processing and the imaging algorithm used to obtain the BINGO maps. We follow the procedure described by \citet{Sazy15} for an earlier version of BINGO.
%Details of the algorithm and data processing can be found in \cite{Sazy15}. 

To assess the reliability with which the cosmological signal can be extracted from the observed data, an end-to-end simulation pipeline has been developed by the BINGO collaboration,
%in Manchester, 
allowing the testing of various aspects of the data analysis pipeline including foreground removal. The input is composed of maps of different emission mechanisms, produced by theoretical models or by observations, as well as by the inherent noise of the instrument and contamination from the environment. Other inputs are features of the telescope, such as the number and arrangement of horns, optical design and receiver characteristics. The pipeline produces as output a time series, which can be turned into maps that simulate the signal picked up by the instrument during a given period of operation.

To detect BAOs in the \textsc{Hi} signal, we will need to remove the contributions from much brighter emission coming from our Galaxy and many extragalactic sources (foregrounds). The most relevant emissions at $\approx$ 1 GHz are a combination of extragalactic point sources and diffuse Galactic synchrotron emission, which taken together is nearly four orders of magnitude larger ($\approx$ 5 K rms at 1 GHz) than the 21 cm signal fluctuations ($\approx$ 200 $\mu$K rms) outside of the Galactic plane. Therefore, the output data are passed through a component separation process to recover the cosmological \textsc{Hi} component.

\begin{figure*}
\begin{center}
        \includegraphics[width=.8\textwidth]{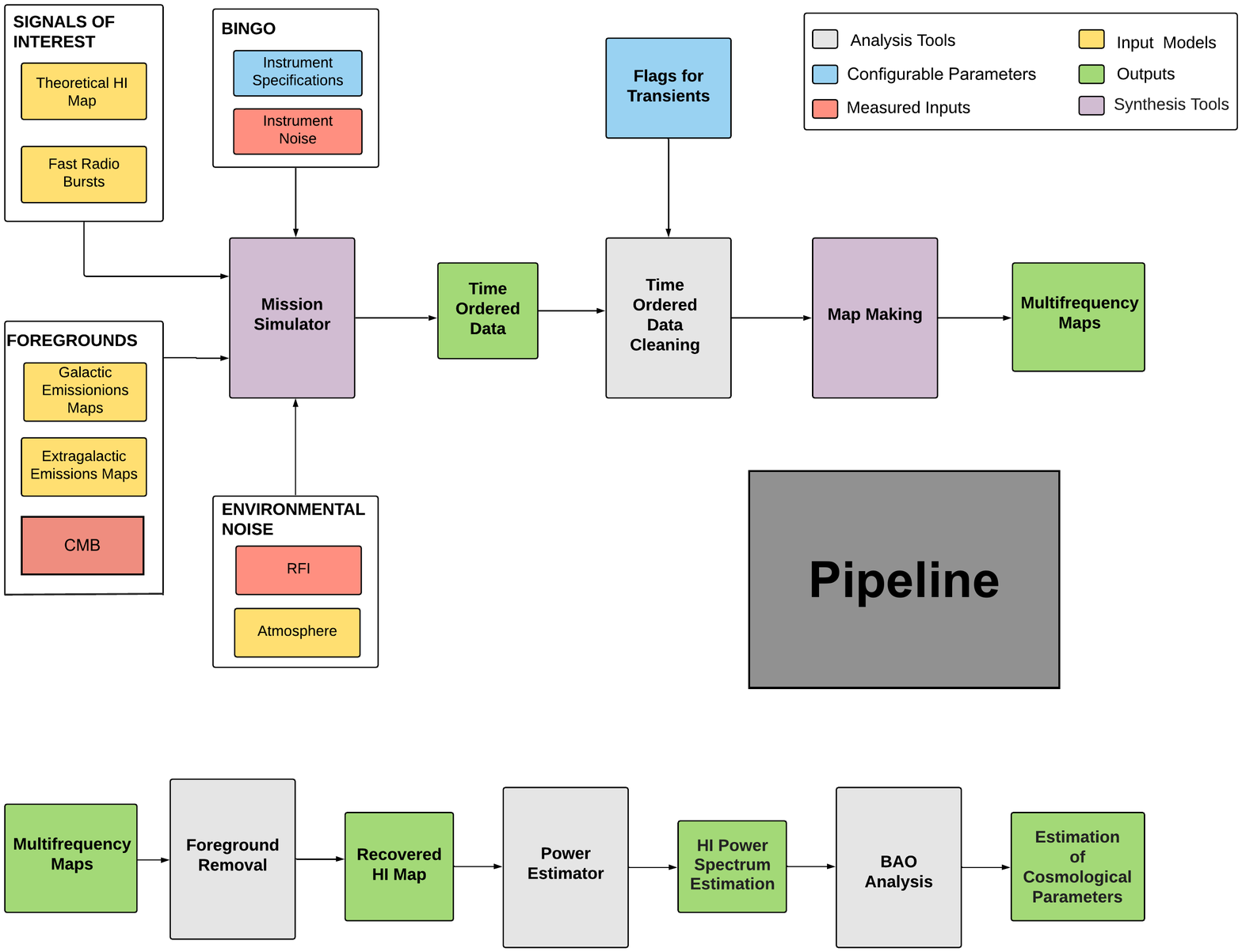}
        \end{center}
    \caption{BINGO mission simulation flowchart. %\RL{I think it's better to remove this ``line'' in the bottom of the figure. It seems it was copied from other place.} \vic{yes, I tried but if you check the figure pdf you can see that the background is perfectly white. I think it is not a big deal, the editorial process will fix that artifact.}
    }
    \label{fig:flow}
\end{figure*}

In Fig. \ref{fig:flow} a flowchart of the BINGO simulation pipeline is shown. In blue are the ``configurable'' parameters of the simulation: instrument specifications (beam shape and number of horns, observation time, knee frequency, number of channels, etc.) and component separation (method and parameters). In red are the parameters that have already been measured (the instrument noise module and the CMB temperature are already included in the pipeline). In yellow are the modules that make use of known models and observations to produce emission maps used as input in the simulations. The RFI part of the pipeline currently only deals with simulating the emissions from the Global Navigation Satellite System (GNSS), and a new module is currently under development to include fixed location terrestrial RFI. 
The atmosphere appears as a source of large-scale black-body emission with a brightness temperature of a few K at around 1\,GHz. 
The atmospheric brightness temperature has been calculated according to the model in \citet{paine19}, and we found, assuming an air temperature of 284\,K, that the value is approximately 4\,K at the zenith.

The pipeline operation makes use of a mission simulator, which processes all the input information described above to produce TODs. In the case of BINGO, this data set consists of the temperature measurements in a given channel $i$ in the range 980--1260\,MHz, in a given celestial coordinate ($\alpha,\delta$) and at a given time of observation $t_{\rm{obs}}$. As BINGO is a telescope without moving parts, each pixel will visit $n$ times the same coordinate ($\alpha$,$\delta$), where $n$ depends on the instrument operating time. Then, for every frequency channel in each horn, the TOD is processed into a map using the {\tt HEALPix} pixelization scheme \citep{Gorski05}. 

%Then, for every horn and frequency channel, the TOD is processed into a {\tt HEALPix} map. 

By default, each pixel in the {\tt HEALPix} map is filled with the mean value of all measurements in that pixel according to equation
\begin{equation}
{T_{\rm{map}}}{\left( {\alpha ,\delta } \right)_{}} = \sum\limits_{t = 0}^t {\left\langle {{T_t}\left( {{\alpha _t},{\delta _t}} \right)} \right\rangle }\,, 
 \end{equation}
where $t$ is the measurement time and $T_{t}$ is the temperature measured at that time in the coordinates ($\alpha$,$\delta$). After this step, which is done separately for each horn, the horn maps are combined into the final cube of BINGO maps.
Eventually, as a separate task from the operation of the pipeline, the set of maps produced by the simulation is processed through a component separation algorithm  to recover the cosmological \textsc{Hi} signal from the data produced by the pipeline.

This work aims to simulate the results to be obtained by the Phase 1 BINGO instrument configuration and to test a new component separation method developed for the Planck mission ({\tt GNILC}), following the results obtained in \cite{Olivari16}. A number of different configurations of the instrument are investigated.

%%%%%%%%%%%%%%%%% NEW SUBSECTION %%%%%%%%%%%%%%%%%%

\subsection{The cosmological signal}
\label{sec:FLASK}

The idealized observed brightness temperature of the sky, $T_{\rm sky}$($\nu$,$\phi$), at frequency $\nu$ and direction $\phi$, is given by %\citep{Wilson13, Condon2016} 
\begin{equation}
\begin{aligned}
T_{\rm{sky}}(\nu ,\phi){\rm{ }} = {T_{\rm{gal}}}(\nu,\phi){\rm{ }} + {T_{\rm{eg}}}(\nu,\phi){\rm{ }} + {T_{\rm{CMB}}}(\nu,\phi){\rm{ }} \\
+ {T_{\rm{atm}}}(\nu,\phi){\rm{ }} + {T_{\rm{COSMO}}}(\nu,\phi)\,,
\end{aligned}
\end{equation}
where $T_{\rm{gal}}$ is the diffuse Galactic radiation, $T_{\rm{eg}}$ is the emission from extragalactic sources, $T_{\rm{CMB}}$($\nu$,$\phi$) is the CMB temperature, $T_{\rm{atm}}$($\nu$,$\phi$) is the atmosphere emission, and $T_{\rm{COSMO}}$ is the cosmological \textsc{Hi} emission. In what follows we briefly describe the cosmological \textsc{Hi} emission model used for the simulations.

The \textsc{Hi} brightness temperature has been simulated with the \textit{Full-sky Lognormal Astro-fields Simulation Kit}  \citep[{\tt FLASK};][]{Xavier16}. {\tt FLASK} can generate fast full-sky simulations of cosmological LSS observables such as multiple matter density tracers (galaxies, quasars, dark matter halos), CMB temperature anisotropies and weak lensing convergence and shear fields. The multiple fields can be generated tomographically in an arbitrary number of redshift slices and all their statistical properties (including cross-correlations) are determined by the angular power spectra supplied as input and the multivariate lognormal (or Gaussian) distribution assumed for the fields. After generating the fields, {\tt FLASK} can apply selection functions and noise to them. 

The \textsc{Hi} emission at low redshifts ($z \lesssim 0.5$, much later than the end of reionization era)
%once the re-ionization period has ended, 
is assumed to be confined to discrete elements such as galaxies, bubbles, and filaments. In this case, the \textsc{Hi} signal is characterized by a mean \textsc{Hi} brightness temperature given by \citep{Hall13}
\begin{equation}
{T_b} = 188{\text{ }}h{\text{ }}{\Omega _{{\textsc{Hi}}}}\frac{{{{\left( {1 + z} \right)}^2}}}{{E\left( z \right)}}{\text{mK}}\,,
\end{equation}
where $E(z)=(H(z)/H_0)$, $H_0=100 h $ km\,s$^{-1}$ Mpc (where $h$ is  the Hubble parameter) and $\Omega _{\textsc{Hi}}$ is the density parameter for \textsc{Hi}. This is the \hi\ assumed value for all papers in this BINGO series.
%where $h$ is the Hubble parameter, $\Omega _{\textsc{Hi}}$ is the \textsc{Hi} comoving mass density in terms of the current critical density (this quantity depends on the \textsc{Hi} luminosity function), $E(z)=(H(z)/H_0(z))$. 
The fluctuations around this mean are due to differences in densities of structure. Once the \textsc{Hi} signal is assumed to be a tracer of the dark matter, these fluctuations can easily be calculated. 
% (typical values in the BINGO redshift range $T_b$ $\sim$ 200 $\mu$K). 

Using the \textsc{Hi} angular power spectra %and assuming Gaussian statistics, 
we produce full-sky maps of the \textsc{Hi} signal with the help of the {\tt synfast} routine of {\tt HEALPix}. However, with the {\tt FLASK} software, we can assume a log-normal distribution for the \textsc{Hi} signal. In doing this, we used the $C_\ell$s from the \textit{Unified Cosmological Library for $C_\ell$s} code \citep[{\tt UCLCL};][]{mcleod2017joint,Loureiro:2018qva} as input for {\tt FLASK}. {\tt UCLCL} is a library for computing two-point angular correlation function of various cosmological fields that are related to LSS surveys. It uses the formalism of angular power two-point correlations and then derives the exact analytical equations for the angular power spectrum of cosmological observables. The auto- and cross-correlations between different observables as well as different galaxy populations (bins) can also be computed.

%The simulated full-sky \textsc{Hi} signal currently includes  30 redshift bins of 21-cm intensity fields for BINGO, equally spaced in frequency. 
The simulated full-sky \textsc{Hi} signal for this work  includes 30 redshift bins of 21 cm intensity fields for BINGO, equally spaced in frequency, following the fiducial BINGO parameters.
The \textsc{Hi} density fields are generated from discrete matter tracers (galaxies) from the DES photometric survey \citep{Flaugher05}. The photo-z distribution for DES galaxies has been estimated from \citet{sanchez14}. The mean temperature of the \textsc{Hi} signal fluctuations in the BINGO redshift range is $\approx$ 200\,$\mu$K. We set {\tt HEALPix} resolution of the map equal to $N_{\rm{side}}$ = 128, which corresponds to a map pixel size of 27 arcmin. 

\begin{figure}
\begin{center}
\includegraphics[width=.48\textwidth]{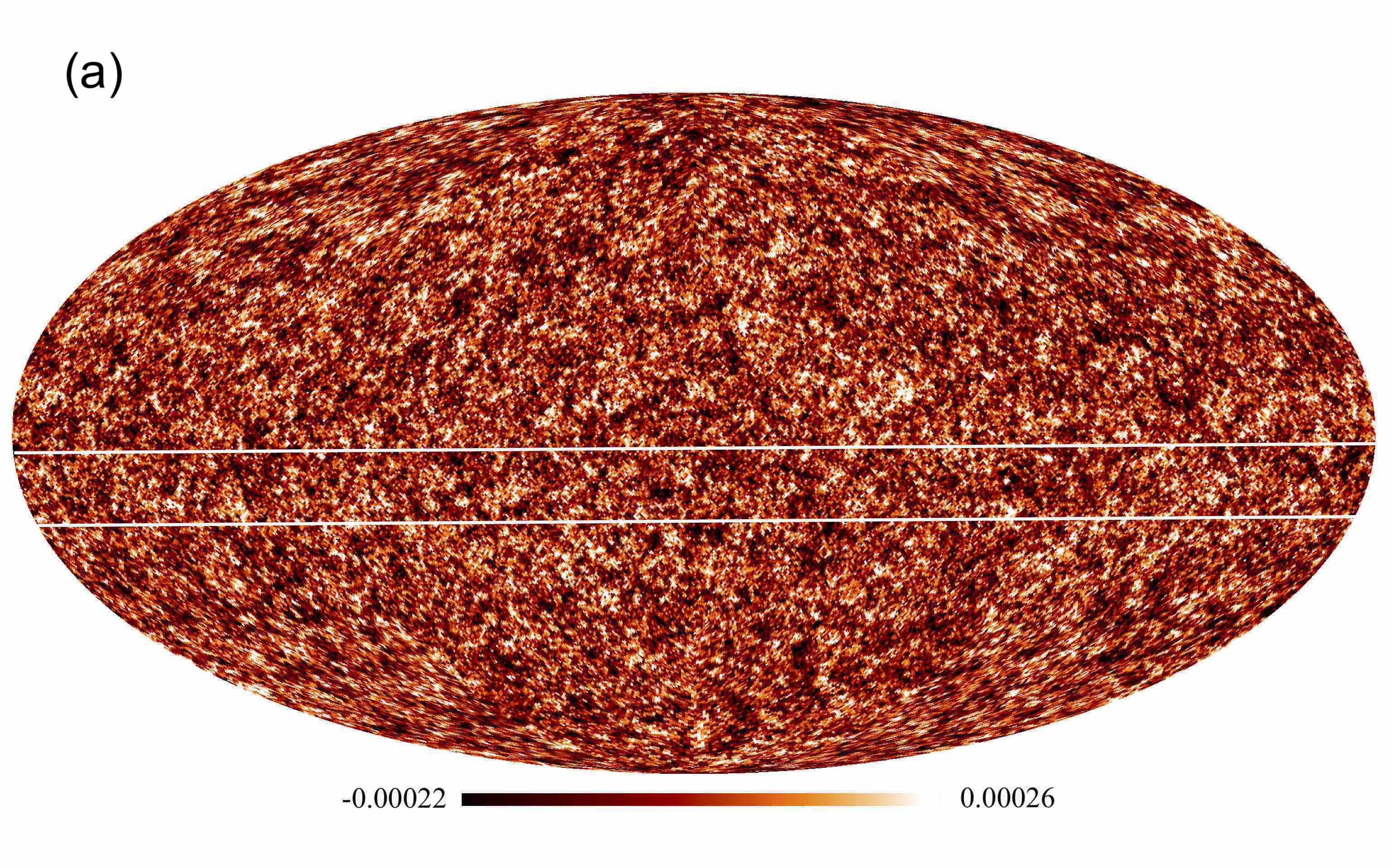}
\includegraphics[width=.48\textwidth]{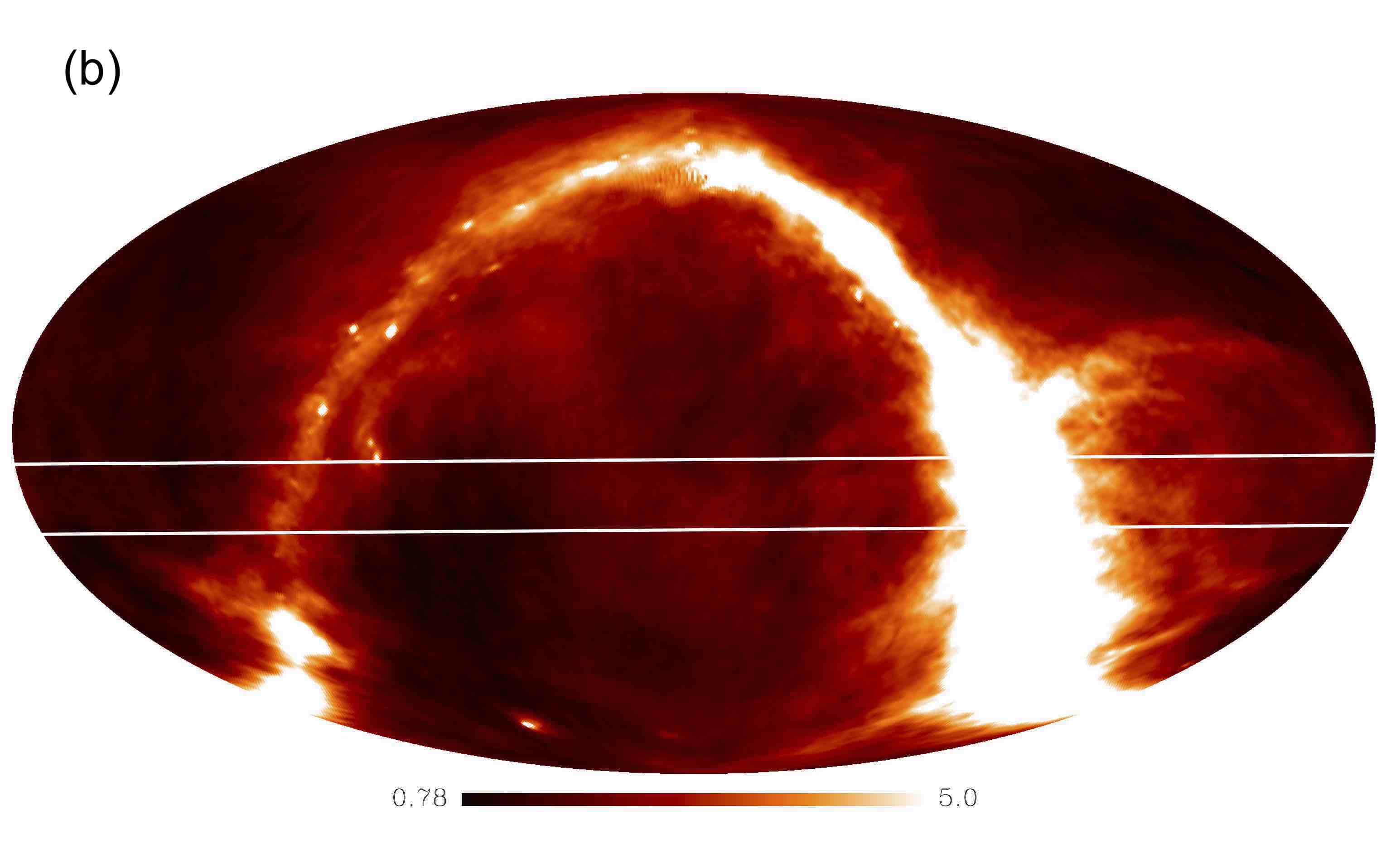}
\includegraphics[width=.48\textwidth]{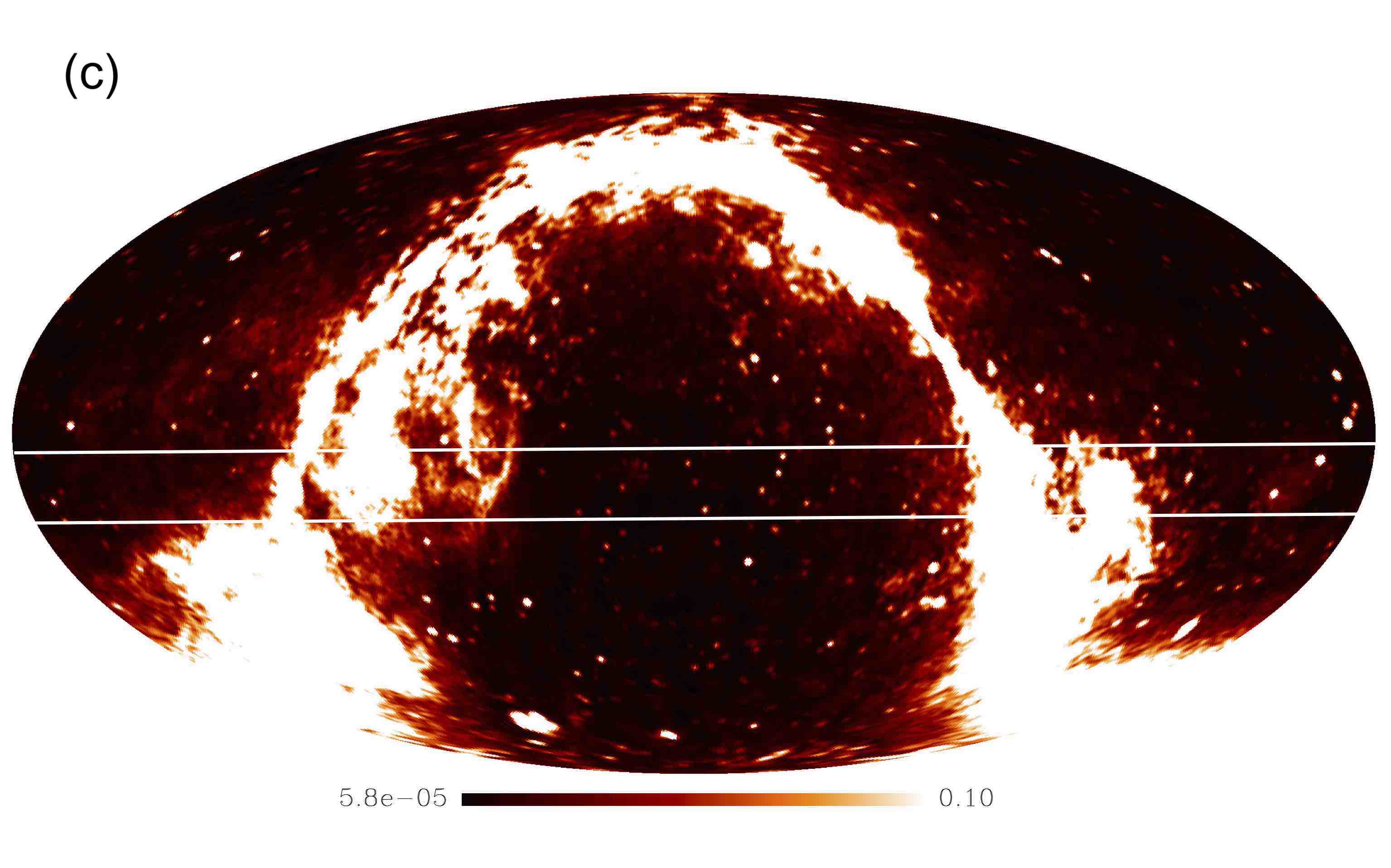}
\includegraphics[width=.48\textwidth]{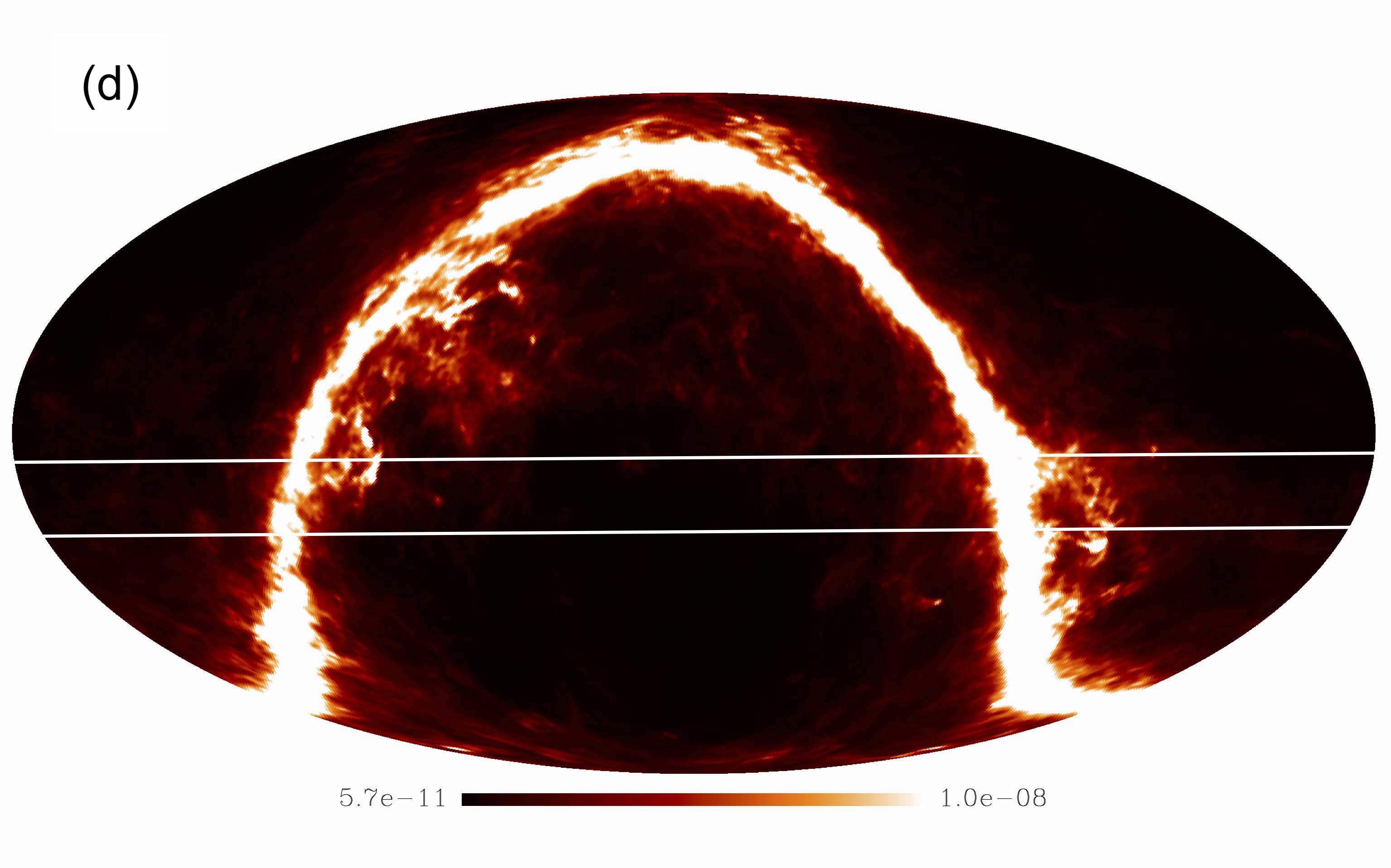}
\end{center}
\caption{Full-sky maps of the cosmological signal and the different foregrounds for a frequency slice $\nu \approx 1.1$ GHz: (a) extragalactic  \textsc{Hi}, (b) synchrotron, (c) free-free, and (d) AME. The stripe defined by the white solid lines is the sky region covered by BINGO. We selected different temperature intervals for the maps to show their features and to allows the comparison of temperature differences at first inspection. Temperatures are given in K.}
 \label{fig:maps}
\end{figure}

In Fig. \ref{fig:maps} the resulting \textsc{Hi} map from the {\tt FLASK} code  at 1.1\,GHz (the central frequency of BINGO bandwidth) is shown. % \RL{are the results similar in other frequencies?} \vic{yes, hereafter we will refer always to the central frequency of bingo}.  
Due to the discrete tracers field that emits the \textsc{Hi} signal, the measured auto-spectra have a shot noise contribution as well as a clustering contribution. To correctly predict the overall \textsc{Hi} signal, this contribution must be accounted for in our \textsc{Hi} simulation. The shot noise depends on the abundance of galaxies observable in our hypothetical survey, and assuming a comoving number density sources $n = \frac{{dN}}{{dV}} = 0.03{h^3}Mp{c^{ - 3}}$, the angular density of the sources can be expressed as
\begin{equation}
\bar N\left( z \right) = 0.03{h^3}\frac{c}{{{H_0}}}\int {{\chi^2}(z)} \frac{{dz}}{{E(s)}}\, .
\end{equation}

Then, the shot noise contribution to the 21 cm power spectrum is subtracted, assuming  assuming a Poisson behavior,
\begin{equation}
C_\ell ^{\rm shot} = \frac{1}{{\bar N\left( z \right)}}\,.
\end{equation}

%%%%%%%%%%%%%%%%% NEW SUBSECTION %%%%%%%%%%%%%%%%%%

\subsection{Foregrounds}
\label{sec:Foregrounds}

An accurate understanding of the foreground emission is essential in order to precisely determine the cosmological  \textsc{Hi} signal. Usually, the principal source of uncertainty is the contamination by foreground emission from the Galaxy, rather than the instrumental noise itself.
Three different types of Galactic foregrounds have been included in the present version of the pipeline: synchrotron, free-free emission and anomalous microwave emission (AME; Fig. \ref{fig:maps}). Extragalactic radio sources, which are an inhomogeneous mix of radio galaxies, quasars and other objects, are being implemented as a separate module in the code.

However, besides the diffuse Galactic emission and the extragalactic radio sources, the CMB also contributes as a contaminant to the \textsc{Hi} signal.  In the BINGO frequency band, the antenna noise temperature and the CMB temperature are of the same order of magnitude ($h\nu /{k_B}T \ll 1$). Therefore, at BINGO frequencies, the CMB radiation represents nearly a constant background (fluctuations of $\approx$ 100 $\mu$K) of 2.7 K. The fluctuations themselves, can be removed by having a spatial CMB template from WMAP/Planck \citep{Planck2018a}, which can then be removed directly from the data (after convolution with the BINGO beam).

For the sake of comparison, in the companion paper V \citep{2020_component_separation} the foreground maps are generated directly from the $Planck$ $Sky$ $Model$ (PSM)  package. They include as input synchrotron, free-free and AME (same as the options in this work) as well as the contribution of a background of radio sources, which we considered less important in a first stage due to the angular scale of our pixelization.

In the following subsections, we briefly describe the specific Galactic foregrounds used in this work and how we simulate them. %following \cite{Olivari16}. 
Further details about the Galactic and extragalactic foregrounds are given in companion paper I \citep{2020_project}. %A slightly different way of generating foregrounds to test component separation methods can be found in the companion paper V \citep{2020_component_separation}.

\subsubsection{Galactic synchrotron}
\label{sec:synchrotron}

It is well known that synchrotron radiation arises from interactions between cosmic ray electrons and magnetic fields in the Galaxy. Since the magnetic fields in our Galaxy extend far to the outskirts of the Galactic plane, synchrotron emission can also be measured at high Galactic latitudes, making it difficult to avoid by only excluding the Galactic plane regions from the analysis. The frequency scaling of synchrotron flux emission is often approximated in the form of a power law, $I_{\nu} \propto \nu^{\epsilon}$, over a limited range of $\nu$. In terms of Rayleigh-Jeans brightness temperature, we have $T \propto \nu^{\gamma}$, with $\gamma  =  - \left( {\epsilon  + 3} \right)/2$. A typical value is $\gamma$\,$\approx -$\,2.5 at radio frequency \citep{Oliveira08}, and takes steeper values $\gamma$ $\approx -$ 3.0 at $\approx$10\,GHz frequencies. Full-sky continuum maps in the low-frequency range are available, for example, the  Haslam  map \citep{Haslam82} at 408\,MHz 
% (50 arcmin angular resolution), 
and the 1.4\,GHz map %(35 arcmin resolution) 
by \cite{Reich86}. 
%In particular, the Haslam map was used as a template to subtract synchrotron foreground in the earlier version of the WMAP analyses.

%Synchrotron emission arises from interactions between cosmic ray electrons and magnetic fields in the Galaxy. The intensity and spectrum depend on the magnetic field strength and cosmic ray energy, and therefore they show significant spatial variations on the sky. For electrons with a power-law energy distribution, $N(E) \propto {E^{ - \alpha }}$, the spectrum of synchrotron emission is 
%
%
%\begin{equation}
%T\left( \nu  \right) \propto {B^{\left( {\alpha  + 1} \right)/2}}{\nu ^\beta },
%\end{equation}
%
%
%
%with $\beta  =  - \left( {\alpha  + 3} \right)/2$ (Rayleigh-Jeans temperature) \citep{Rybicki79}. 

\begin{figure}
    \centering
        \includegraphics[width=\columnwidth]{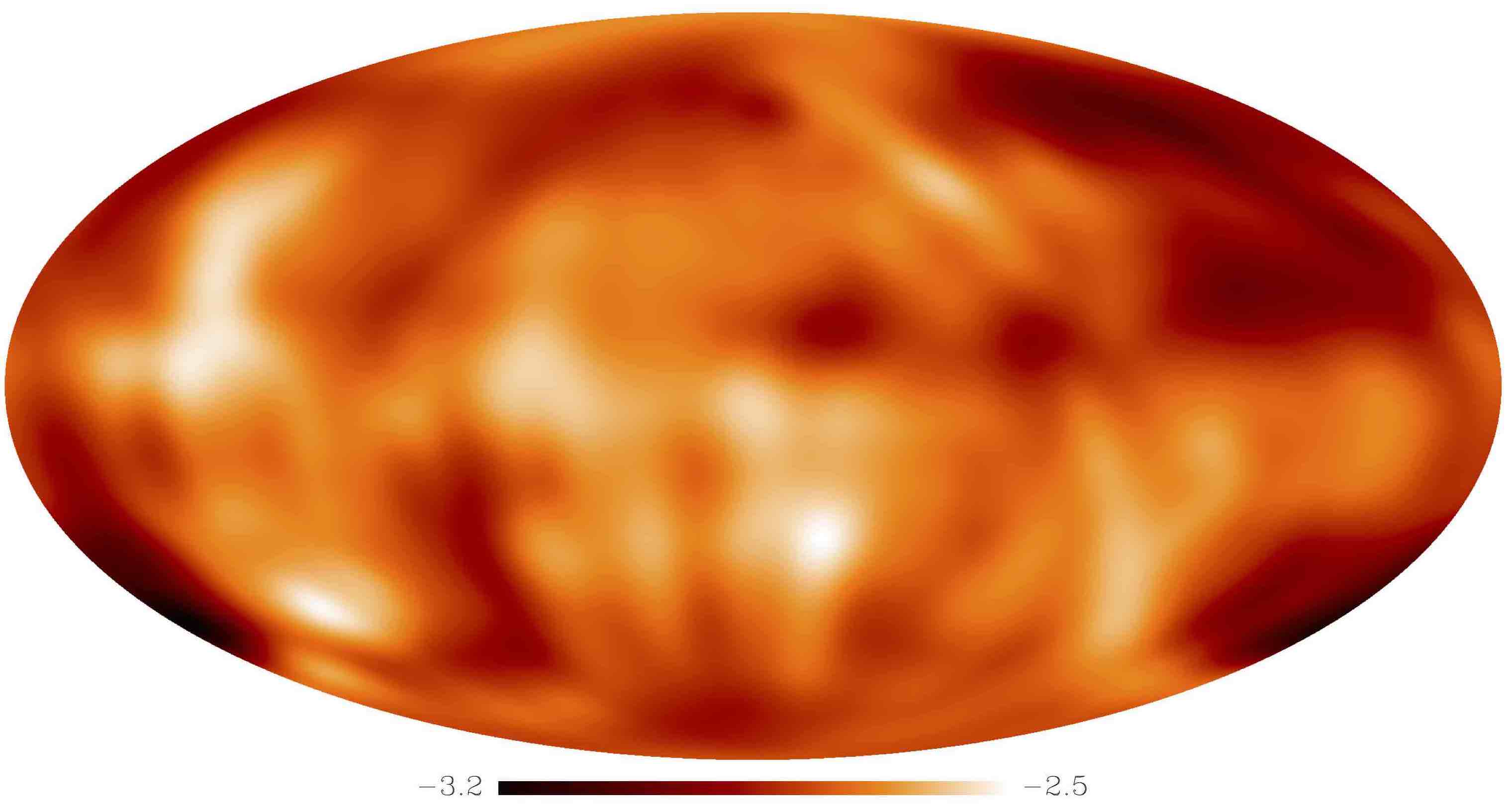}
    \caption{Map of the synchrotron spectral index according to the Giardino model.}
    \label{fig:synch_index}
\end{figure}

For our simulations, we used the reprocessed Haslam 408 MHz all-sky map from \cite{Remazeilles15}. This includes artificially added small-scale fluctuations as described by \citet{Delabrouille13}. We consider that the synchrotron spectral index is spatially variable according to the Giardino model \citep{Giardino02}, which was derived using the full-sky map of synchrotron emission at 408\,MHz from \cite{Haslam82}, the northern hemisphere map at 1420\,MHz from \cite{Reich86} and the southern hemisphere map at 2326\,MHz from \cite{Jonas98}. In the Giardino model, the synchrotron spectral index has a mean value of $-$ 2.9 and a standard deviation of 0.1 (Fig. \ref{fig:synch_index}) and is good for frequencies $\lesssim$ 2.3\,GHz. Our choice here differs from companion paper V \citep{2020_component_separation}, who use the PSM synchrotron sky with the synchrotron spectral index distribution produced by \cite{Miville08}.

The observed diffuse synchrotron emission at radio frequencies is distributed across the entire sky as shown in the point-source subtracted all-sky map in Fig. \ref{fig:maps}. Most of the synchrotron emission from the Galaxy is concentrated along the Galactic plane, but large-scale features such as Loop I stretch over around half of the north Galactic sky.

%The synchrotron emission can be highly polarized, up to 50 - 70\%, with higher values at higher galactic latitudes \citep{Kogut07}.
%The larger polarization degrees at high Galactic latitudes are mostly attributed to the local structures, namely, the Fan region and the North Galactic Spur, which have polarization degrees as large as $\sim$ 30\%.
%Accurate modeling of the Galactic cosmic ray and magnetic field distributions can in principle be used to predict the polarization foreground from synchrotron emission and remove it from observed maps.

\subsubsection{Free-free}
\label{sec:free}

%It is possible to calculate the brightness temperature at frequency $\nu$ for free-free emission,
%
%
%\begin{equation}
%{T_{ff}} \simeq 0.082{T_e}^{ - {{0.35}_\nu } - \alpha }\int\limits_{l.o.s.} {{N_e}{N_i}dl}, 
%\end{equation}
%
%
%where $N_{e}$ is the electron density, $N_{i}$ is the ions density,  $T_{e}$ is the electron temperature,  $dl$ is the infinitesimal element of distance along the line of sight and  $\alpha$ is the spectral index varying from 2.1 to 2.15 with errors $\pm$0.03 \citep{Delabrouille07}. 

%Free-free radiation is not polarized because it is an incoherent emission from individual electrons scattered by nuclei in a partially ionised medium.
% Interaction of free electrons with ions in ionised media makes the electrons lose energy by emitting photons. 
Since, in the BINGO frequency interval, free-free radio continuum emission is subdominant to other Galactic emission types such as synchrotron emission it is very challenging to uniquely isolate a map of radio free-free emission. Therefore, generating free-free emission maps of the whole sky mostly relies upon the use of tracer emissions. However, free-free maps can also be produced by component separation techniques (e.g., \citealp{pla_2016}), which could be used here in the future.

At optical wavelengths, the emission from the H$\alpha$ transition can be used as a tracer of free-free \citep{Dickinson03}. Optical H$\alpha$ continuum maps can be easily related to free-free emission at radio wavelengths in regions with a small H$\alpha$ optical depth ($\tau$ $<$ 1), which is limited to the sky far from the Galactic plane. We use the H$\alpha$ map by \cite{Dickinson03} as a template for the Galactic free-free emission. This map includes small-scale fluctuations as described in \cite{Delabrouille13}.
The free-free spectrum can be well defined by a power law with a temperature spectral index $\beta=-$ 2.1 \citep{Dickinson03},
\begin{equation}
{T_{\rm{f\,f}}} \approx 10\,{\rm{mK}\,}{\left( {\frac{{{T_e}}}{{{{10}^4}{\,\rm{K}\,}}}} \right)^{0.667}}{\left( {\frac{\nu }{{{\rm{GHz}}}}} \right)^{ - 2.1}}\left( {\frac{{{I_{H\alpha }}}}{{\rm{R}}}} \right)\,,
\label{eq:free}
\end{equation}
which flattens the spectral index of the total continuum of our Galaxy where the free-free has a brightness temperature comparable to that of the synchrotron emission.
$T_{e}$ is the electron temperature in K, and $I_{H \alpha}$ is the H$_{\alpha}$ template whose emission is given in Rayleigh (R). 

In the BINGO observation region, the free-free brightness temperature fluctuations are $\approx$ 0.25\,mK; therefore, they are considerably weaker than the synchrotron component but significantly brighter than the \textsc{Hi} fluctuations \citep{Battye13}. The final free-free template at 1.1\,GHz is shown in Fig. \ref{fig:maps}.

\subsubsection{Anomalous microwave emission}
\label{sec:AME}

Anomalous microwave emission is diffuse Galactic radiation detected at frequencies between 10\,GHz and 60\,GHz \citep{Dickinson18}. The best accepted physical process for the production of AME is the spinning dust model, which proposes that the rapid rotation of the electric dipoles associated with the smallest dust grains in the interstellar medium can generate microwave frequency emission that peaks between 10 and 60\,GHz.
%Far infrared and submillimeter observations showed that the thermal emission spectrum of dust can be represented by a modified blackbody function, which depends on the spectral index, dust emissivity and equilibrium temperature of the interstellar medium \citep{Tibbs12}.

%The spatial distribution of the anomalous emission and its frequency dependence have been estimated by subtracting contributions of synchrotron, free-free, and thermal dust emissions (e.g., assuming a constant spectral index $\alpha$) from the WMAP 23, 33, 41 and 61 GHz maps \citep{Miville08}. Observations from the \textit{Planck} satellite in recent years have provided a wealth of new information on spinning dust emission. \citep{Planck2016} derived an all-sky map of spinning dust emission using the Gibbs-sampling component separation algorithm Commander \citep{Eriksen08}, which revealed direct correlations between spinning dust and several other Galactic components at high Galactic latitudes.

\begin{figure}
    \centering
        \includegraphics[width=\columnwidth]{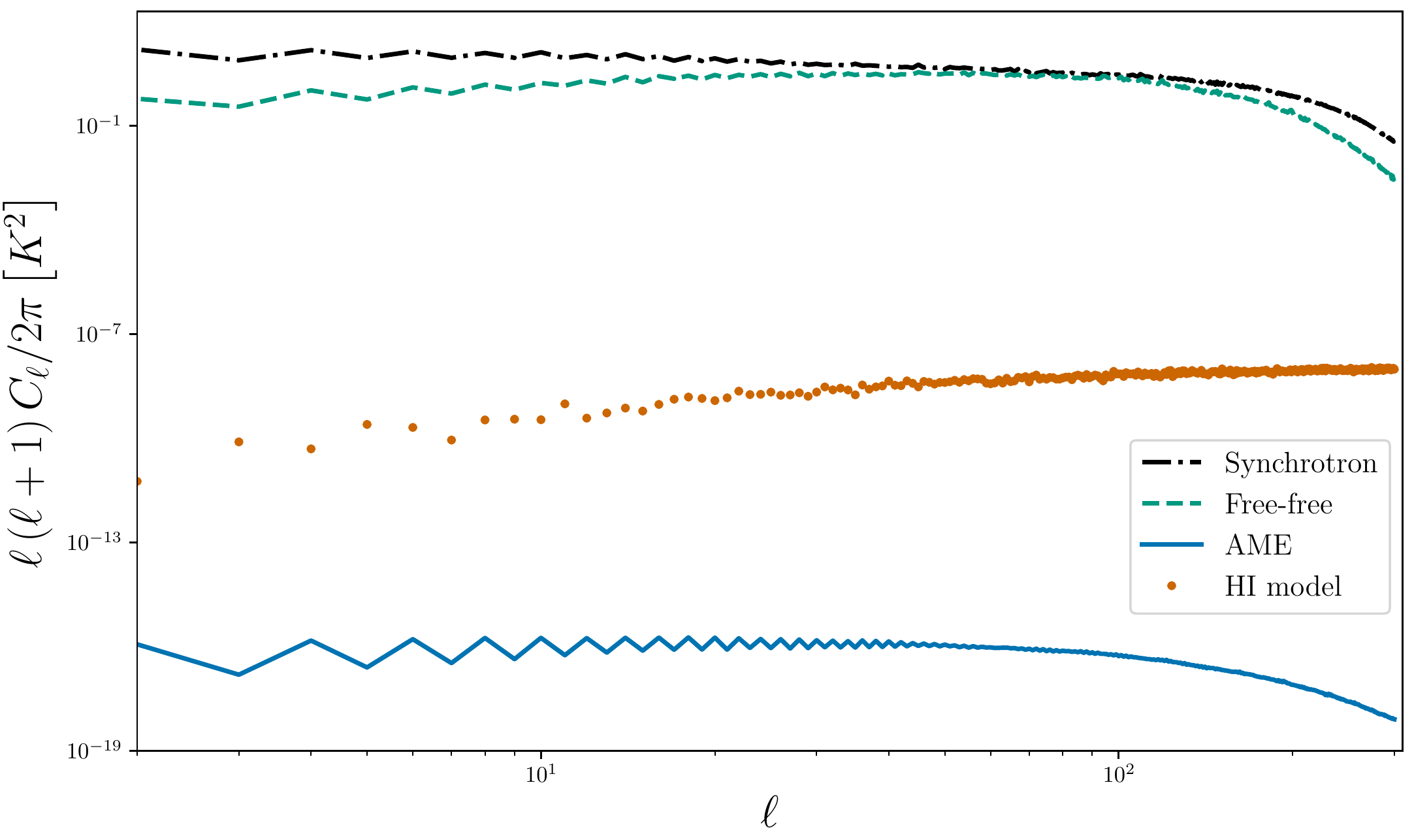}
    \caption{Angular power spectra of the cosmological signal and the different foreground components at $\nu$ $\approx$ 1.1 GHz.}
    \label{fig:foregrounds}
\end{figure}

To simulate the AME emission, we used as a template the \textit{Planck} $\tau_{353}$ optical depth map \citep{PlanckXLVIII}. We adopted the factor 8.3 $\times$ 10$^{6}$ $\mu$K/$\tau_{353}$ to convert the dust optical depth at 353 GHz to the AME temperature at 22.8\,GHz, in units of $\mu$K \citep{Planck2016}. To scale the AME emission from 22.8\,GHz to the BINGO frequencies of $\approx$ 1\,GHz, we used the publicly available {\tt spdust2} code \citep{Silsbee11}, which calculates the spinning dust emissivity as a function of frequency for various environments of the interstellar medium.
Anomalous microwave emission temperature fluctuations are extremely weak ($\approx$  2\,$\mu$K) at frequencies below 10\,GHz and are negligible in the frequency range  of BINGO (Fig. \ref{fig:foregrounds}), but we included them for completeness. The final AME template at 1.1\,GHz is shown in Fig. \ref{fig:maps}.

\subsection{Instrumental noise}
\label{sec:noise}

\subsubsection{Thermal noise}

%The current concept of the telescope gives a larger field-of-view and the optimization of the number of horns improves the overall sensitivity of the experiment.
Phase 1 of BINGO will provide a large field-of-view, and the fine-tuning in the positioning of the 28 horns will improve the overall sensitivity of the experiment. The parameters that affect the sensitivity are the survey area $\Omega_{\rm{sur}}$, the beam size $\theta_{\rm{FWHM}}$, the number of horns $n_{f}$, and the integration time $t_{\rm{obs}}$.
The r.m.s. noise per pixel is given by the well-known radiometer equation \citep[see][]{Wilson13}:

\begin{equation}
        \sigma_t = \frac{T_{\rm sys}}{\sqrt{t_{\rm pix}\delta \nu }}    \,,
        \label{eq:noise}
    \end{equation}with $\delta {\nu}$ as the frequency bin width and $t_{pix}$ the observation time per pixel related to the total observing time by

\begin{equation}
{t_{\rm{pix}}} = {t_{\rm{obs}}}{n_f}\frac{{{\Omega _{\rm{pix}}}}}{{{\Omega _{\rm{sur}}}}}\,,
\label{eq:noise2}
\end{equation}where $\Omega_{\rm{pix}}$ is the pixel solid angle.
Equation \ref{eq:noise2} shows that increasing the field-of-view results in higher r.m.s noise, whereas adding more horns reduces it. The compromise option is determined by the balance between the cosmic variance and systematic effects that dominates the error at large scales and the thermal error dominated at small scales. The cosmic variance error can be reduced with a larger sky coverage. The thermal noise amplitude of the instrument $\sigma _{t}$ has been calculated according to the number of horns (Eq. \ref{eq:noise2}). Table \ref{thermal_noise} presents the estimated sensitivities for the scenarios considered in this study, as well as values for other cases.

\begin{figure}
    \centering
        \includegraphics[width=\columnwidth]{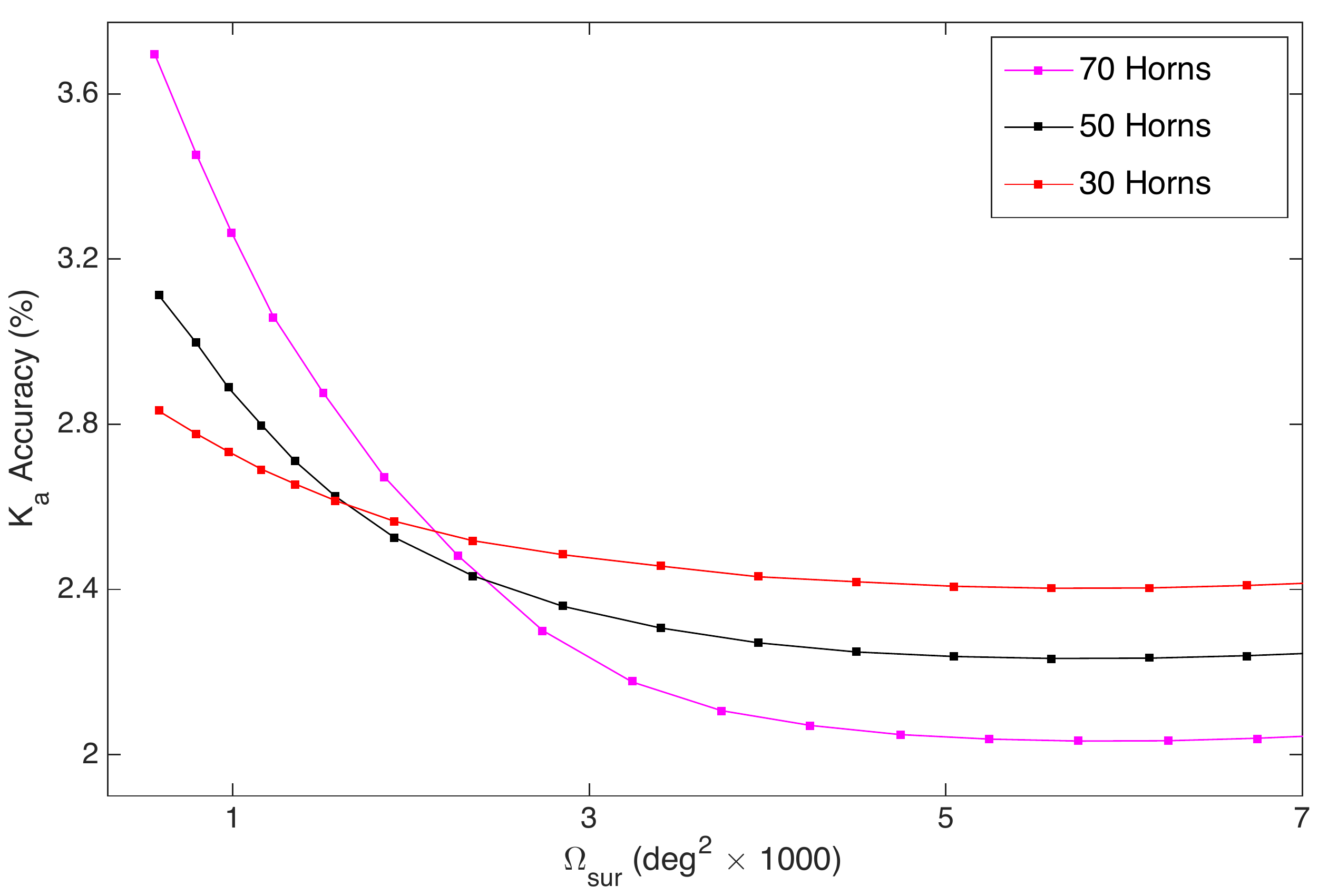}
    \caption{Uncertainty on the acoustic scale for 30, 50, and 70 feeds based on a one-year observation time and $\theta_{\rm{FWHM}}$ = 40 arcmin.}
    \label{fig:accuracy}
\end{figure}

In theory, in order to improve the constraints on the acoustic scale $k_{A}$, we require a larger field-of-view. However, a larger focal plane area likely brings with it more systematic errors. The criteria is the uncertainty on the acoustic scale $k_{A}$, related to the constraints on the measurements of the BAO features. Following the analysis carried out in \citet{Battye13}, if we consider one central redshift $z$ = 0.3 it is possible to probe the acoustic scale $k_{A}$ with a fractional error of 2.4\% with 2000 deg$^{2}$ and 50 horns. We find that increasing the number of horns from 50 to 70 and the survey area from 2000 deg$^{2}$ to 4000 deg$^{2}$  gives a measurement of the acoustic scale $k_{A}$ with accuracy 2.0\% (Fig. \ref{fig:accuracy}). In order to determine the optimal concept, there is a balance to find between the cosmic variance and $1/f$ noise that dominates the error at large scales and the thermal error dominated at small scales \citep{Blake_2003, Seo10}.
The latter can be reduced with larger field-of-view and number of horns. At large scales the limitation on the accuracy on the measured power spectrum due to the sample variance can be improved by larger survey volume, and so more sky coverage. However, we have to take into account the systematics induced by a larger focal plane area. Optical simulations show that going further away from the center of the focal plane induces a slightly loss in terms of performance of the beam (ellipticity, gain). 
In this case, the configuration with $\approx$ 50 horns and 5000 deg$^{2}$ field-of-view represents a good compromise ($\delta$ $ k_{A}$/$k_{A}$ $\approx$ 2.1\%). During Phase 1, BINGO will operate with 28 horns and we intend to add 28 more in Phase 2. 

%We note that for the sky coverage of about 5000 deg$^2$, even for 28 horns the difference in accuracy between 28 and 56 is slightly less than 0.25\%. The improvement in sensitivity versus the cost of a focal plane with more than 56 horns is not justified.
%The extra cost of a larger focal plane does not justify the improved sensitivity.

\begin{table}
\centering
\caption{Thermal noise amplitude for different numbers of horns (considering a 100\% duty cycle).}
%\vspace{2pc}
\resizebox{9cm}{!}{
\begin{tabular}{c|c|c} 
\toprule
{\bf Array - Number of feeds}   & {\bf Integration time (yr)} & {\bf $\sigma_{t}$ ($\mu$K)}    \\ \midrule \midrule
Double-rectangular - {\bf 28}  & 1 (2)    & 30 (21) \\ \midrule 
Hexagonal - {\bf 31}  & 1 (2)    & 29 (20) \\ \midrule
Double-rectangular - {\bf 56}  & 1 (2)    & 15 (11) \\ \midrule
Hexagonal - {\bf 49}  & 1 (2)    & 16 (11) \\ \bottomrule
\bottomrule
\hline
\end{tabular}
}
\label{thermal_noise}
\end{table}

%\begin{table}
%\centering
%\caption{Thermal noise amplitude for each observational scenario.}
%\vspace{2pc}
%\resizebox{8cm}{!}{
%\begin{tabular}{c||c||c||c} \toprule
%{\bf Number of feeds}       & 30    & 40        & 50        \\  \midrule       
%{\bf Integration time (y)}  &  1    & 1         & 1 \space \space \space \space \space 2 \\ \midrule   
%{\bf $\sigma _{t}$ (mK)}    & 0.04  &  0.035    & 0.032 \space \space \space \space 0.022  \\ %\bottomrule
%    \bottomrule
%\end{tabular}
%}
%\label{thermal_noise}
%\end{table}

\subsubsection{$1/f$ noise}

The term $1/f$ noise refers to the type of noise with a power spectrum density (PSD) of the form $S(f)$ $\propto$ $1/f^{\alpha}$ where the spectral index $\alpha$ identifies specific types such as pink noise \mbox{($\alpha$ = 1)} and red noise ($\alpha$ = 2). In most cases, $\alpha$ is equal to 1. The frequency taken into account for the noise is the inverse of the observation time ($f$ = 1/$t_{\rm{obs}}$). Even though it is widely observed in a broad range of scientific areas, a fundamental description of $1/f$ noise is yet to be found. %In the case of BINGO, the $1/f$ noise is produced by gain fluctuations of the amplifiers and is, therefore, correlated across all frequency channels.
In the case of BINGO, the $1/f$ noise is produced by gain fluctuations of the amplifiers and therefore it is expected to be strongly correlated across all frequency channels.

It is common in astronomy to define the PSD of a receiver contaminated with  thermal and $1/f$ noise as follows: for a system with knee frequency $f_{k}$ (defined as the frequency to which thermal and $1/f$ noise have the same PSD value), system temperature $T_{\rm{sys}}$, the PSD of the noise fluctuations is given by a power-law model \citep{Harper18b}

\begin{equation}
S\left( {f,\omega } \right) = \frac{{T_{\rm{sys}}^2}}{{\delta \nu }}\left[ {1 + C\left( {\beta ,{n_{ch} }} \right){{\left( {\frac{{{f_k}}}{f}} \right)}^\alpha }{{\left( {\frac{1}{{\omega \Delta \nu }}} \right)}^{\frac{{1 - \beta }}{\beta }}}} \right]\,,
\label{eq:psd}
\end{equation}
where $\omega$ is the Fourier mode of the spectroscopic frequency $\nu$, $\Delta\nu$ is the total bandwidth of the receiver and $C$($\beta$,$N_{\nu}$) a normalization factor given by ($N_{\nu} - 1)/(2N_{\nu}\delta{\nu}$), $\delta{\nu}$ is the channel bandwidth and $\alpha$ is the spectral index of the correlated noise. The spectral index $\beta$  describes a $1/f$ noise identical in every receiver channel ($\beta$ $\approx$ 0) and a noise that is independent in every channel \mbox{($\beta$ = 1)}. For an ideal receiver, $f_{k}$ should be 0 meaning
that the receiver TOD is dominated by flat power-spectrum thermal (white) noise only.  The thermal and $1/f$ noise are both simulated from independent noise realizations. The stability of the BINGO noise properties can be quantified by the variation in the
white noise, knee frequency and spectral index over the lifetime of the experiment.
The presence of $1/f$ noise in an observation map introduces stripes following the scan circle strategy \citep{Maino02}, and its main effect is to increase the uncertainty of measurements on large spatial scales. This striped structure appears because the mean level of the noise is, in general, different for each circle of measurements, as shown by \citet{Janssen96}. It is important to note the difference to the companion paper V approach, where the details of the instrument observation strategy are not considered. Their analysis does include the $1/f$ component but not the $\beta$ factor, which accounts for correlations across the frequency channels.

In these simulations, the $1/f$ noise fluctuations are assumed to be small multiplicative variations around the system temperature and Gaussian distributed. The 1/f noise can be represented in the TOD as

\begin{equation}
\Delta T\left( {t,\nu } \right) = \delta G\left( {t,\nu } \right){T_{\rm{sys}}}\left( {t,\nu } \right),
\end{equation}
where $\Delta T\left( {t,\nu } \right)$ is the power of the $1/f$ fluctuations at time $t$ and frequency $\nu$, which is the combination of the instantaneous fluctuation in the gain $\delta{G}$ and system temperature $T_{\rm{sys}}$. The PSD of $\delta{G}$ can be described by two parts. The first part is the power spectrum of the temporal fluctuations as in Eq. \ref{eq:psd} but without the thermal noise component

\begin{equation}
P(f)=\frac 1 {\delta \nu}\left(\frac {f_k}f\right)^ \alpha,  
\end{equation}
while the second component of the $1/f$ power spectrum describes the correlations of the noise in frequency, and may be described by a conservative power-law model

\begin{equation}
F\left( \omega  \right) = {\left( {\frac{{{\omega _0}}}{\omega }} \right)^{\frac{{1 - \beta }}{\beta }}},  
\end{equation}
where $\omega$ is the Fourier mode of the spectral frequency (i.e., the wavenumber), $\omega_{0}$ is the smallest wavenumber (1/$\Delta{\nu}$), and $\beta$ describes the frequency correlation. The simulated gain fluctuations should be interpreted as ripples across the 2D observed region.

\begin{figure}
    \centering
        \includegraphics[width=\columnwidth]{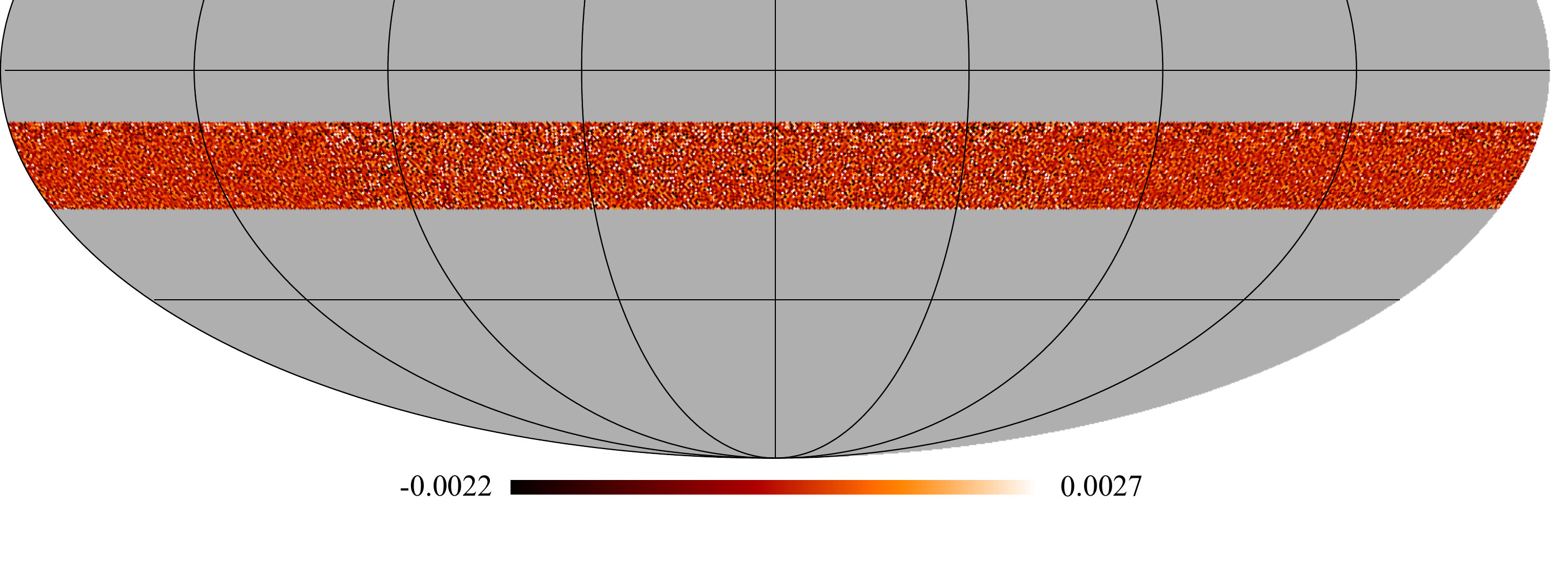}
    \caption{Simulated $1/f$ noise map of the BINGO observational region. The striped structure is reduced after the filtering of the data. The scale quantifies fluctuations $\delta T$ around the average in K.}
    \label{fig:1/f}
\end{figure}

 In Fig. \ref{fig:1/f} we show a $1/f$ map (there is no thermal noise) of the BINGO region using the \mbox{28 horn} double-rectangular array. When $\beta$ = 1, the frequency spectrum is entirely uncorrelated. This means that the number of modes needed to describe the $\beta$ = 1 $1/f$ noise is equal to the number of channels; therefore, removing the $1/f$ noise will be very challenging for typical component separation methods (see Sect. \ref{sec:results_p}). For the simulation in Fig. \ref{fig:1/f} we assumed a spectral index $\beta$ = 0.25 and the same value of $f_{k}$ = 1\,mHz for each receiver and for a 9.33\,MHz channel bandwidth. The index $\alpha$ is assumed to be 1, while $\beta$ and $f_{k}$ were obtained from preliminary measurements, at Jodrell Bank Observatory, of the statistical properties of the noise using data from the BINGO test correlation receiver \citep{Evans2017}. The $1/f$ noise has been filtered on timescales of $\approx$ 6 minutes, which is the time structures with the angular scales of interest take to drift across the field-of-view of one horn. This assumes that $1/f$ noise is fully calibrated out (e.g., using a calibration diode) on such timescales, and represents an optimistic scenario.

A good front end for a radio telescope designed for IM with the required stability will exhibit a low knee frequency,  ideally on the order of one thousandth of a hertz. The use of correlation receivers using colfets as reference sources in the BINGO telescope will allow for efficient accounting of the $1/f$ noise contribution \citep[see][]{2020_instrument}. To reduce the stripe contamination in the maps, we need to carefully measure the relative gains of the individual receivers and determine the beam contribution for each horn before combining the signals from all the beams. 
%In order to produce maps without stripes we will need to measure the relative gains of the individual beams before combining the signals from all the beams.

%%%%%%%%%%%%%%%%% NEW SUBSECTION %%%%%%%%%%%%%%%%%%

\section{Observational strategy}
\label{sec:strategy}

%\subsection{Observational approach}
%\label{sec:approach}

In the radio band the natural tracer  of LSSs is the 21 cm line of \textsc{Hi}, but the volume emissivity associated with this line is low, meaning that detecting individual galaxies at z $\approx$ 1 requires a very substantial collecting area. Interferometer arrays are likely to be the best approach to probing higher redshifts at z $\approx$ 1, where an angular resolution of $\approx$ 0.1$^{\circ}$ is required. Using a single dish, moderate-sized telescope with an ultra-stable receiver system, is the lowest-cost approach to IM measurements of BAOs at low redshifts \citep{Battye13} and intermediate angular scales.  The idea is to exploit the broad beam at low-frequencies to carry out IM \citep{Peterson06, Masui13} and consequently measure the overall integrated \textsc{Hi} brightness temperature of a large number of galaxies, taken together as a tracer of the LSS.
Detecting signals of $\approx$ 200 $\mu$K with a non-cryogenic receiver of standard performance implies that every pixel in our intensity map requires an accumulated integration time of $>$ one day over the course of the observing campaign. The total integration time can be built up by many returns to the same patch of sky, but between these, the receiver gains need to be stable. %Achieving this is a major design consideration as well as a major operation concern. 
%The relative strength of the foregrounds,
The Galactic emission is known to be significantly polarized. The synchrotron emission, for instance, is polarized up to 40-50\% \citep{Wolleben06}. The linearly polarized part of the Galactic emission will undergo Faraday rotation as it travels through the Galactic magnetic field and the interstellar medium.
This means that the observations need to be made with clean beam(s) with low side-lobe levels and very good polarization purity in order to add foreground degrees of freedom.

For the following analysis, we chose a declination strip that minimizes the contribution from Galactic foregrounds. We assume that the telescope will map a $\approx$ 15$^{\circ}$ declination strip centered at $\delta=$ -17.5$^{\circ}$ as the sky drifts past the telescope.
The need to resolve structures of angular sizes corresponding to a linear scale of around 150 Mpc in our chosen redshift range implies that the required angular resolution has to be about 40 arcmin.
As the Earth rotates, each BINGO horn observes a ring at a single declination set up by the instrument geometry. One complete ring is scanned in 24 hours, which is the periodicity of the sky signal. Therefore, a set of periodic rings (one per horn) with 24 hours each, is a standard representation of the BINGO data. The arrangement of feed horns in the focal plane has been optimized in such a way as to cover the $\approx$ 15$^{\circ}$ declination strip and at the same time to have some redundancy, that is to say, beams have some superposition with beams of adjacent horns (as shown in Fig. \ref{fig:arranjo_28}). This will increase the signal to noise ratio.

% \begin{figure}[h]
%     \centering
%       \includegraphics[width=0.9\columnwidth]{figures/double_all_sky}
%     \caption{All-sky hits map of the BINGO drift scan strip. The observed sky region has been simulated using the double-rectangular array and integrating for one year.}
%     \label{fig:all_sky}
% \end{figure}

%The arrangement of feed horns in the focal plane has been optimized in such a way to cover the $\sim$ 15$^{\circ}$ declination strip and at the same time to have some redundancy, i.e.,  the same area of the sky is observed by multiple horns, in order to increase the signal to noise ratio. 

The positioning of each horn is defined by two parameters. The first is the Cartesian ($x$,$y$) coordinates of the horn in the focal plane and the second is the elevation and azimuthal angles (el, az). Horns located in the outermost regions of the array are slightly tilted with respect to the focal plane to under-illuminate the secondary mirror and to reduce sidelobes and ground contaminations. The details of the BINGO optical design are given in \citet{2020_optical_design}. %In this work we assume that the horns are arranged in the hexagonal (Fig. \ref{fig:arranjo_31}) and the double-rectangular (Fig. \ref{fig:arranjo_28}) array formats. The beams are given by a circular Gaussian and are diffraction-limited. 
In this work, we assume two different horn configurations from that paper: the hexagonal (Fig. \ref{fig:arranjo_31}) and the double-rectangular (Fig. \ref{fig:arranjo_28}) formats. The beams are well approximated by a Gaussian shape and are diffraction-limited.

Different observation times have been tested to understand how they affect the \textsc{Hi} signal recovery. The uniformity of the sky coverage will depend on the pixel size used. %{\color{red} \Large Therefore in order to ensure uniform coverage of the sky for BINGO, where the sky will drift by the focal plane, there should be N independent sight-lines when the resolution of each pixel in the final map is $\sim$ 15$^{\circ}$/N.}
Since our observation method relies upon the sky drifting across the focal plane, we guarantee a complete (and uniform) sky coverage for BINGO allowing for N independent lines-of-sight, with the resolution of each pixel in the final map being $\approx$ 15$^{\circ}$/N.
Both arrangements  are good enough when mapping the sky using {\tt HEALPix} with  $N_{\rm{side}}$ = 64 (larger pixels). This $N_{\rm{side}}$ roughly corresponds to pixels of size 54 arcmin, which avoids missing pixels. This means the effective beam of BINGO will be broader, but by about 30\% However, gaps will appear in the sky coverage (declination direction) when the resolution is increased to the real BINGO resolution $\theta_{\rm{FWHM}}$ = 40 arcmin. 

%Figure \ref{fig:nside} shows the resulting coverage when using the arrangement shown in Figure \ref{fig:arranjo} (31 horns). In all the sky coverage maps, x-axis is right ascension centred around our Galaxy visible within the BINGO strip, and the y-axis is declination centred around -18$^{\circ}$. The gaps in the sky coverage are self evident for $Nside$ = 128 and 256.

Other arrangements have been analyzed like the ones from \citet{Battye13} and \citet{Sazy15} or with 60 horns placed equidistantly along the vertical axis, and in such a way as to cover $\approx$ 15$^{\circ}$ in total.  
However, the sky coverage obtained with the above-mentioned options and a resolution $N_{\rm{side}}$ = 128 (27' pixels) leaves gaps between data stripes, with unobserved regions at constant declination.
%{\color{red} \Large However, the coverage results are incomplete along the declination axis for 27 arcmin pixels ($N_{\rm{side}}$ = 128).} 
There are three possible ways of overcoming the problem. Only use {\tt HEALPix} with maximum $N_{\rm{side}}$ = 64 to produce the maps, change the declination coverage by varying the pointing of the full focal plane, or build a focal plane with more horns. The last option is not possible for financial reasons. Changing the declination coverage can be accomplished by moving all horns vertically at different steps. Five different horn vertical positions are allowed inside the 2400\,mm tall hexagonal case. These vertical displacements should happen in such a way as to minimize declination separation between adjacent horns, meaning vertical positions change in the focal plane as a whole (all horns should be placed in the same  ‘‘new'' position), as opposed to elevation and azimuth displacements, which may occur for individual units. Maximum displacement is $\pm$ 300\,mm for the focal plane.

\begin{figure}[t]
    \centering
        \includegraphics[width=0.93\columnwidth]{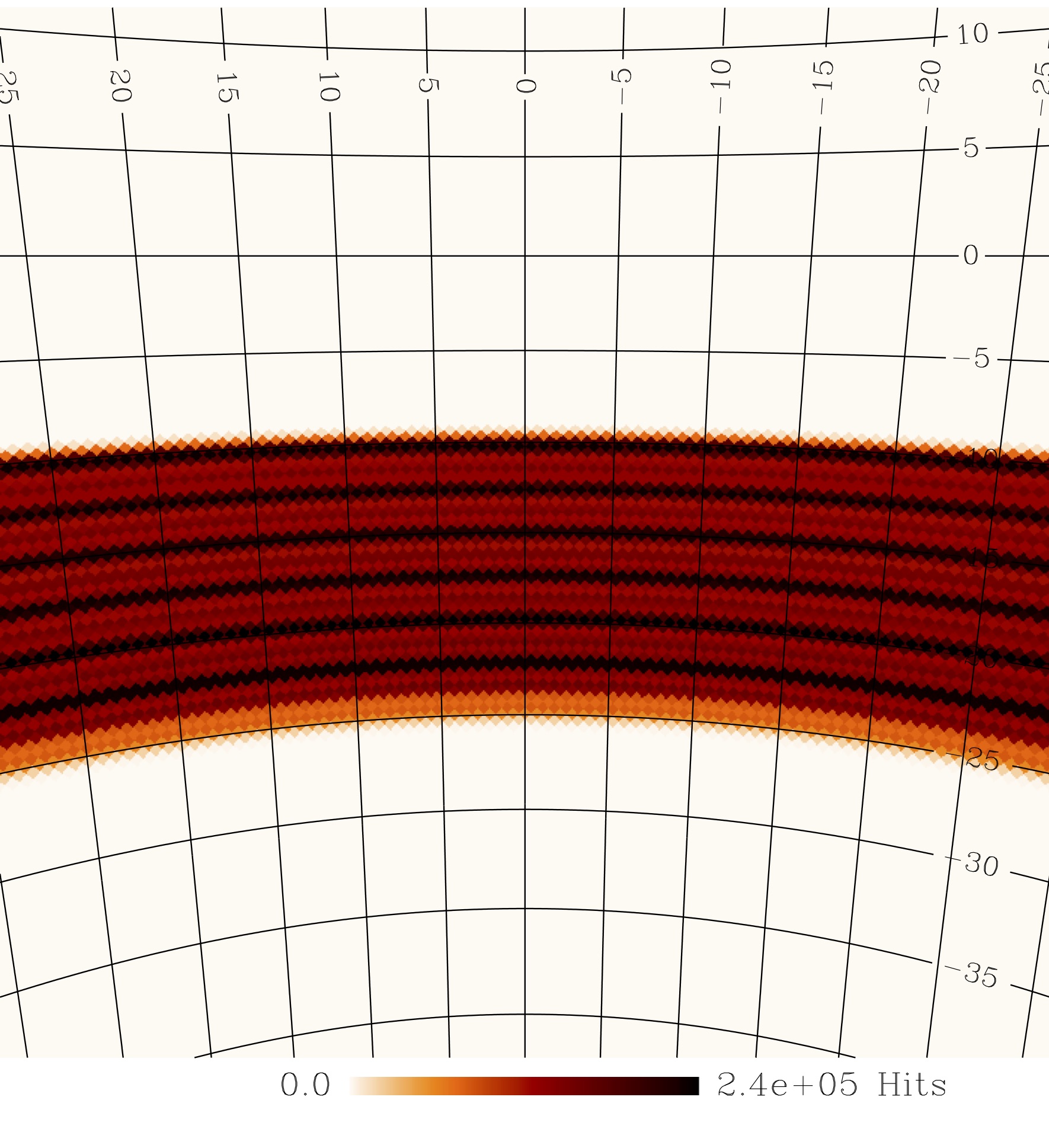}
    \caption{Gnomonic projection centered at $\delta$ = -17.5$^{\circ}$, RA = 0 of the BINGO sky coverage when using the hexagonal arrangement with 31 feed horns, after 5 years of mission.}
    \label{fig:hexagonal_sum}
\end{figure}

\begin{figure}[t]
    \centering
        \includegraphics[width=0.93\columnwidth]{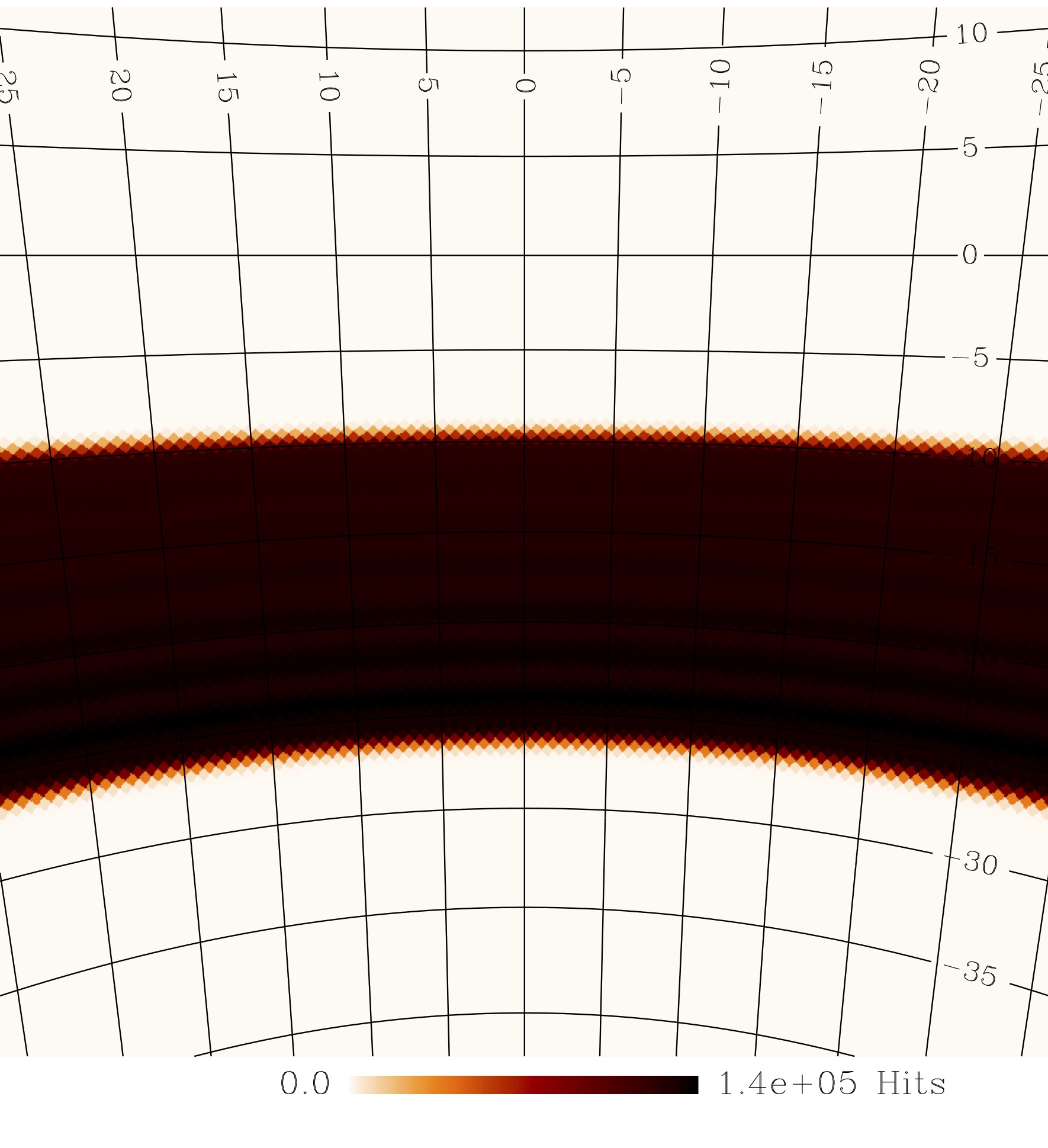}
    \caption{Gnomonic projection centered at $\delta$ = -17.5$^{\circ}$, RA = 0 of the BINGO sky coverage when using the double-rectangular arrangement with 28 feed horns, after 5 years of mission. }
    \label{fig:double_sum}
\end{figure}

We simulated the resulting BINGO coverage after displacing all the horns up and down in the focal plane every year by $\pm$ 150\,mm steps. In doing this, we generated hits maps (meaning how many times a given pixel is scanned) relative to the central channel, which corresponds to the frequency interval 1110.67--1120.00\,MHz. All maps are created with $N_{\rm{side}}$ = 128 and then degraded to BINGO resolution ($\theta_{\rm{FWHM}}$ = 40 arcmin). The ``five elevation'' summed maps are equivalent to five years of observations.
The resulting maps are shown in Fig. \ref{fig:hexagonal_sum} and Fig. \ref{fig:double_sum}. The horn repositioning strategy results in a more homogeneous covered area (we found 25\% more coverage than the fixed elevation option for the hexagonal format and 12\% more coverage for the double-rectangular), which will allow us to obtain a more uniform signal-to-noise ratio per pixel and a better recovery of the overall \textsc{Hi} signal. Regarding the two arrangements, we found differences in terms of sky coverage and uniformity. The declination strip is $\approx$ 15$^{\circ}$ for the hexagonal array, whereas in the case of the double-rectangular is $\approx$ 17.5$^{\circ}$ ($\approx$ 17\% larger).

\begin{figure*}[ht]
\begin{center}
\includegraphics[width=.85\textwidth]{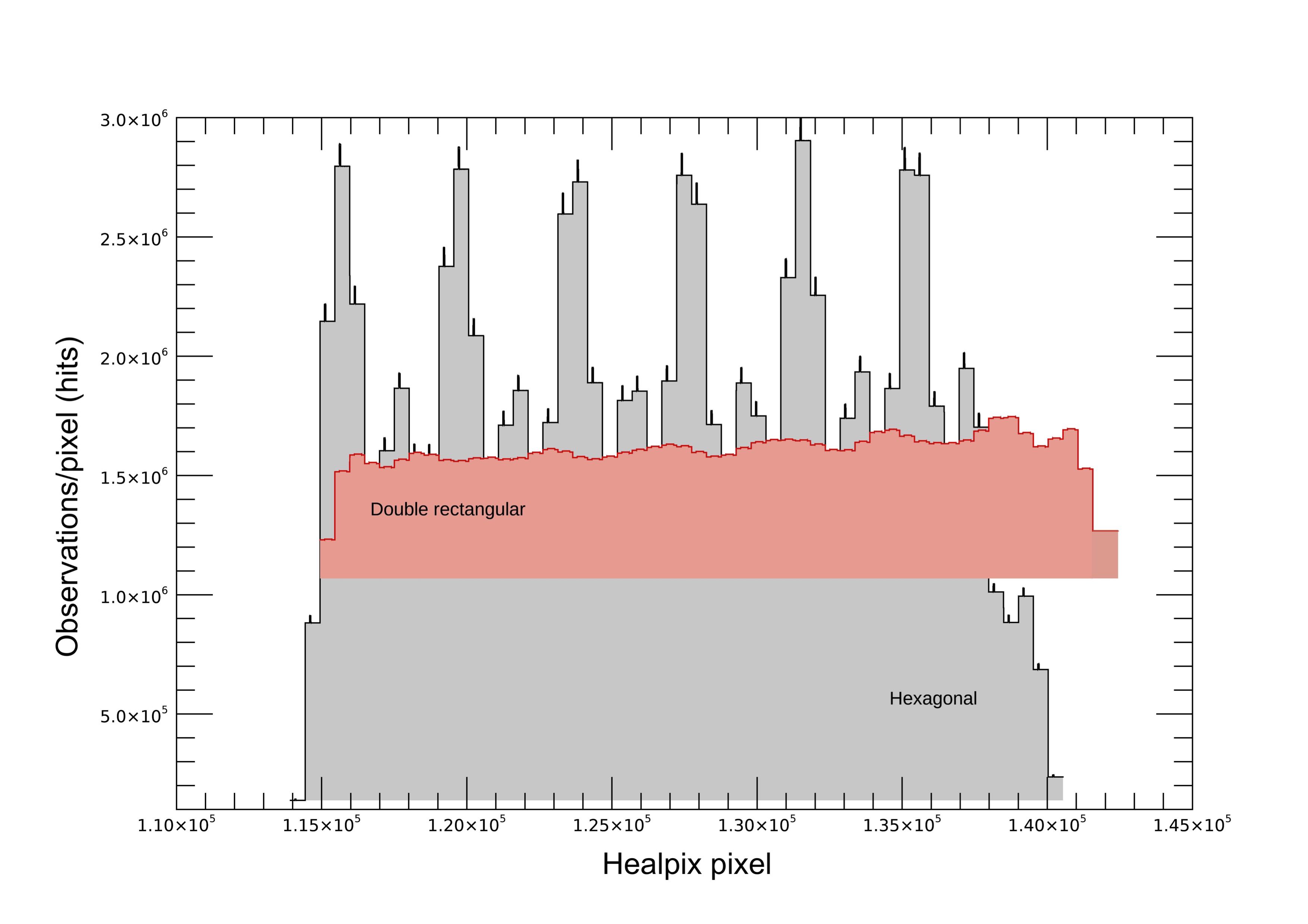}
\end{center}
\caption{Sky coverage for BINGO observing strategy (pixel size = 40 arcmin). The histograms represent the sky coverage obtained with the hexagonal ($gray$) and the double-rectangular ($red$) arrangement. The histograms have been obtained with five different horn positionings ($\pm$ 300\,mm, $\pm$\,150 mm and 0) considering one year of integration for each elevation. On the $y$ axis we have the number of observations relative to each pixel of the map.}
 \label{fig:coverage}
\end{figure*}

The better sky coverage uniformity achieved with the double-rectangular array is clearly visible in Fig. \ref{fig:coverage}, where the minimum number of observations per pixel (hits) as a function of the covered area ({\tt HEALPix} pixel) is shown. The peaks of the hexagonal arrangement are due to feed redundancy along the scan direction (i.e., pixels seeing the same sky each day) since there are more horns aligned in cross-elevation. %Although this observational strategy allow us to avoid gaps in the sky coverage, is clearly visible how different feed arrays can affect the uniformity. Better results might be obtained with larger steps but the actual geometry of the supporting structure prevents this option.
It is clearly visible how different feed arrays can affect the uniformity in the sky coverage. The strategy of displacing the horns vertically, reconfiguring the focal plane, allow us to avoid gaps in the sky coverage. Better results might be obtained with larger vertical displacements, but the actual geometry of the supporting structure prevents this option.

In terms of noise power spectrum, there is additionally an impact when the observation time of the $N$ horns is not uniformly spread over a fraction of the sky. It is worth noting that the noise power spectrum of an homogeneous coverage is independent of the pixel size used to produce the map.
In the case of inhomogeneous coverage, the noise power in a spherical harmonic transform (computed from the harmonic coefficients of the noise map, excluding pixels that are not observed) is
\begin{equation}
{N_\ell } = 4\pi  \times \left\langle {\sigma _t^2} \right\rangle /{N_{\rm{pix}}} = \frac{{4\pi }}{{{N_{\rm{pix}}}}} \times \sum\limits_p {\left[ {\frac{{\sigma _{\rm{noise}}^2}}{{{\tau _{\rm{obs}}}\left( p \right)}}} \right]}\,,
 \label{eq:noisepow}
\end{equation}
where $N_{\rm{pix}}$ is the total number of pixels in the map, $\tau_{\rm{obs}}(p)$ is the total time spent observing pixel $p$ (summing-up the time of observation by all horns) and $\sigma_{\rm{noise}}$ is the white noise level

\begin{equation}
{\sigma _{\rm{noise}}} =  \frac{{{T_{\rm{sys}}}}}{{\sqrt {\delta \nu } }}\,.
 \label{eq:white}
\end{equation}

Hence, the amplitude of the noise power spectrum increases when the time distribution is not uniform. When the inhomogeneity is not too large, such an increase remains small. We found a noise level $\approx$ 7\% greater when the sky is scanned with the hexagonal horn configuration compared to the double-rectangular arrangement and $N_{\rm{side}}$ = 128. However, as the $N_{\rm{side}}$ of the maps is increased, big gaps appear and the noise level can reach values of $\approx$ 70\% (in the extreme case where half of the pixels are observed five times more often than the other half).

In the double-rectangular configuration, two rows of detectors are shifted compared to the original first two rows of detectors by a one-quarter height of the hexagon height, whereas in the hexagonal configuration the difference in horn heights between the first and the second columns represents half the hexagonal height. This indicates we can reach a better than Nyquist configuration (i.e., having a number of samples per beam equal to two) with the double-rectangular array by simply shifting the position of the horns once during the survey lifetime. Since the shifting occurs each year of the survey, we can obtain a map that is over-sampled in the $y$ direction compared to Nyquist sampling. This will allow us to use other techniques to extract further resolution from the maps such as the drizzle technique applied to {\tt HEALPix} maps \citep{Paradis12}. Finally, we can say the double-rectangular array give a better noise distribution, which is good for the quality of the data that can be expected from BINGO, but reduces the redundancy.

%\textcolor{cyan}{\Large To be added by FILIPE- Furthermore, this focal plane configuration will ensure a sufficient Nyquist sampling per beam-width ($\sim$ 1.2) during the observing campaign.}

%%%%%%%%%%%%%%%%% NEW SUBSECTION %%%%%%%%%%%%%%%%%%

\section{Component separation}
\label{sec:gnilc}
\subsection{GNILC}

The signal measured by a radio telescope is the composition of cosmological signals emitted in the early Universe (e.g., CMB or cosmological \textsc{Hi} signal), astrophysical sources emitting in the late Universe (e.g., Galactic foregrounds and extragalactic point sources) and systematic noise of the instrument (e.g., thermal and $1/f$ noise). A component separation process aims to extract the signal of interest from the measured signal by evaluating the correlations of the measurements at different frequencies using physical emission models. This process is extremely important for \textsc{Hi} IM experiments, since the detected signal is typically smaller than the Galactic foreground contribution by a factor of roughly $10^{-4}$. In addition to these components, systematic contributions can also be removed during the separation process. 

%This process is extremely important for IM experiments, since the detected signal is dominated by foregrounds. These can be about $10^{4}$ times larger than the signal to be recovered, as in the case of  \textsc{Hi} emission. In addition to these components, systematic contributions can also be removed with the separation process. 

Several component separation techniques are available in the literature concerning foreground removal, in particular for CMB data. For instance, the principal component analysis (PCA) technique was successfully employed in the detection of \textsc{Hi} at $z$ = 0.8 by cross-correlation using the Green Bank 100 m Telescope \citep{Switzer13}.
The Generalized Needlet Internal Linear Combination \citep[{\tt GNILC};][]{Remazeilles11}, is a component separation method developed for the \textit{Planck} collaboration and applied to IM experiments by \cite{Olivari16}. For this method, a data set containing the measurements of intensity (or temperature) $x_{i}(p)$ at a given frequency $i$ and at a given pixel $p$ can be represented by
\begin{equation}
{x_i}\left( p \right) = {s_i}\left( p \right) + {n_i}\left( p \right),
 \label{eq:gnilc}
\end{equation}
where $s_{i}(p)$ is the map of the cosmological signal to be recovered and $n_{i}(p)$ is the map of foregrounds emission and instrumental noise.
Equation \ref{eq:gnilc} can be also written as a 1 $\times$ $n_{ch}$ vector, where the $n_{ch}$ is the number of channels (frequency bins)
\begin{equation}
{\textbf{x}}\left( p \right) = {\textbf{s}}\left( p \right) + {\textbf{n}}\left( p \right)\,.
\label{eq:gnilc2}
\end{equation}

In this work what is assumed to be the noise are the foregrounds plus the $1/f$ component. \textsc{Hi} and thermal noise are about the same order of magnitude in some multipole scales and so, as we try to recover \textsc{Hi}, some thermal noise will be recovered as
well. In the {\tt GNILC} method a set of windows (``needlets'') is defined in harmonic space so that specific different ranges of
angular scales of the input map are isolated. Needlets are a particular construction of wavelets family that can be interpreted as
band-pass filters, $h_l^{\left( j \right)}$, in harmonic space and can be defined such that
\begin{equation}
\sum\limits_j^{} {\left[ {h_l^{\left( j \right)}} \right]}  = 1\,,
\end{equation}
where $j$ defines an interval of multipoles. The $a_{lm}$ harmonic coefficients are filtered and transformed back into a map, so the statistical information contained there is only relative to a certain range of angular scales. It means that, for each frequency map, there are several needlet maps. This step permits a more localized analysis of the signal. Next, the data covariance matrix  \textbf{R}, whose dimensions are $n_{ch}$ $\times$ $n_{ch}$, is defined at pixel $p$ by
\begin{equation}
\textbf{R}\left( p \right) = {\textbf{R}_\textsc{Hi}}\left( p \right) + {\textbf{R}_n}\left( p \right)\,,
\end{equation}
where ${\textbf{R}_\textsc{Hi}}\left( p \right)$ = $\langle$${\textbf{s}}\left( p \right)$ ${\textbf{s}^{T}}\left( p \right)$$\rangle$ is the \textsc{Hi} covariance matrix, and ${\textbf{R}_n}\left( p \right)$ = $\langle$${\textbf{n}}\left( p \right)$ ${\textbf{n}^{T}}\left( p \right)$$\rangle$ is the covariance matrix of the noise (foregrounds plus $1/f$). The number of components representing the observation data is limited to the number of channels of the experiment, $n_{ch}$. The foreground components are frequency correlated, so that the foreground signal plus noise \textbf{n} can be represented by a linear combination of $m$ independent vectors.

The signal to be recovered can be estimated by
\begin{equation}
{\rm{\hat s}} = \textbf{W}\textbf{x}\,,
\end{equation}
where \textbf{W} is the $n_{ch}$ $\times$ $n_{ch}$ weight matrix and \textbf{x} the data vector described in Eq. \ref{eq:gnilc2}. The matrix \textbf{W} minimizes the total variance of the estimated vector ${\rm{\hat s}}$, under the condition \textbf{W}\textbf{S} = \textbf{S}, so that
\begin{equation}
{\rm{\textbf{W} = \textbf{S}}}{\left( {{{\rm{\textbf{S}}}^T}{{\rm{\textbf{R}}}^{ - 1}}{\rm{\textbf{S}}}} \right)^{ - 1}}{{\rm{\textbf{S}}}^T}{{\rm{\textbf{R}}}^{ - 1}}\,,
\label{eq:gnilc3}
\end{equation}
where \textbf{S} is the estimate \textsc{Hi} plus noise mixing matrix. In order to use Eq. \ref{eq:gnilc3} to recover the signal of interest, it is necessary to estimate the \textbf{S} matrix. To achieve this aim, a theoretical power spectrum of \textsc{Hi}  is used to determine the local ratio between the cosmological signal and the total observed signal. 
At this point, we have an estimate for the reconstructed signal ${\rm{\hat s}}$ for each needlet scale. Finally, for each frequency channel, the needlet maps are added to give a complete {\tt GNILC} recovered  \textsc{Hi} plus thermal noise map.

\subsection{Results}
\label{sec:results_p}

Most of the %Manchester 
pipeline is written in the {\tt Python} programming language. Maps are generated using the {\tt HEALPix} pixelization scheme. The code simulates the $1/f$ noise-contaminated TODs used for generating maps. Time-ordered
data sets with correlated $1/f$ noise properties result in images containing stripes along the telescope drift directions that can dominate the astronomical signal. A set of simulations has been used to test the performance of the BINGO telescope and the quality of the component separation method.

\begin{table}
\centering
\caption{Simulation parameters.} 
%\vspace{2pc}
\resizebox{7cm}{!}{
\begin{tabular}{c||c} \toprule 
    {\bf Parameter}                 & {\bf Value} \\ \midrule \midrule             
    Beam resolution ($^{\circ}$)    &  0.67 \\ \midrule 
    Observation time (yr)           & 1, 2  \\ \midrule 
    Frequency range (MHz)           & 980 - 1260 \\ \midrule 
    Number of feeds $n_{f}$         & 28, 56 \\ \midrule 
    Number of channels $n_{ch}$     & 30 \\ \midrule 
    %Survey area (deg$^{2}$)         & $ \approx $ 6070 \\ \midrule  
    Knee frequency (Hz)             & 0.001 \\ \midrule 
    $1/f$ spectral index $\beta$    & 0.001, 0.12, 0.25, 0.6 \\ \midrule 
    $T_{\textsc{sys}}$ (K)          &  70 \\   \bottomrule
    \bottomrule
\end{tabular}
}
\label{parameters_simul}
\end{table}

The {\tt GNILC} method is also investigated in the BINGO paper V \citep{2020_component_separation} with contaminants generated from the PSM code. In this paper, the efficiency of {\tt GNILC} in reconstructing \textsc{Hi} plus thermal noise maps in the presence of contaminants (already described in this work) is analyzed in respect of how the change of $1/f$ parameters, number of feed horns and observation time influence this process. 
%The efficiency of GNILC in reconstructing \textsc{Hi} plus thermal noise map in the presence of 1/f noise and the foregrounds already described will be analyzed, but with attention on how the change of 1/f parameters, number of feed horns and observation time influence this process.
To estimate the dimension of the \textsc{Hi} plus noise subspace in its PCA step, {\tt GNILC} makes use of theoretically known \textsc{Hi} plus noise power spectra (or \textsc{Hi} plus noise template maps). The reason for using the \textsc{Hi} signal plus the instrumental thermal noise as the signal of interest is that these two emissions, for most of the current IM experiments, are roughly of the same order of magnitude for some of the scales of interest (smaller scales or higher multipoles). 
Therefore, even when we try to recover the \textsc{Hi} signal alone, we end up recovering some thermal noise as well at these scales. This, however, is not an optimal reconstruction of the \textsc{Hi} plus noise signal, since {\tt GNILC} will try to remove as much noise as possible from the data. To avoid creating artificial artifacts on the noise maps, the most efficient strategy for \textsc{Hi} IM is then to recover both the \textsc{Hi} and noise signals as one single component.

%For the analysis a GAL070 $Plank$ HFI Galactic mask was used, with a cosine apodization of width 3 to avoid boundary artefacts in the power spectrum estimation. All maps are generated with $Nside$ = 128. Afterwards, the maps are convoluted with the telescope beam size $\theta_{FWHM}$. 

We used a galaxy mask, similar to the GAL70 Planck \textsc{Hi} mask, with a cosine apodization of $3^{\circ}$ to avoid boundary artifacts in the power spectrum estimation. All maps are generated with $N_{\rm{side}}$ = 128. Table \ref{parameters_simul} shows the instrumental parameters used for the simulations. Some values have been modified in each tested scenario, such as the observation time, the number of feed horns and the $1/f$ spectral index. The number of channels has been limited to $n_{ch}$ = 30. This is a compromise between the increase in the thermal noise amplitude (Eq. \ref{eq:noise}), computational processing time and the improvement in the {\tt GNILC} performance with an increase in the number of channels.

\begin{figure}
    \centering
        \includegraphics[width=0.9\columnwidth]{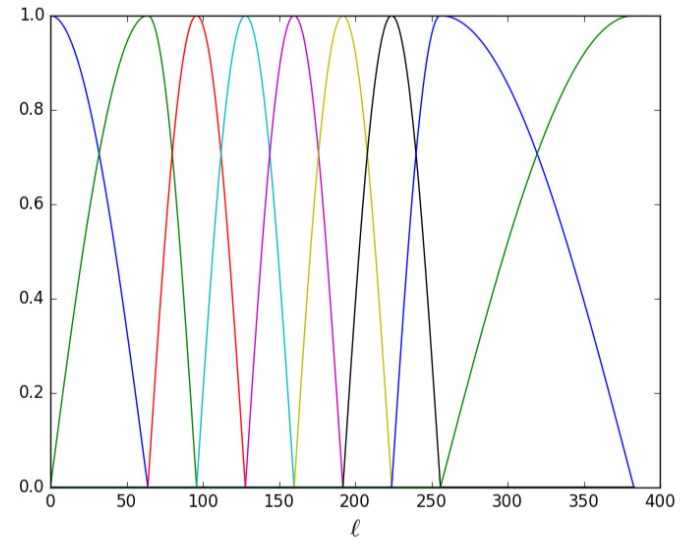}
    \caption{Particular set of needlets used in this work. A needlet is a tool that permits a certain range of multipoles (or physical scales) to be isolated for the benefit of a particular analysis or procedure.}
    \label{fig:needlet}
\end{figure}

The {\tt GNILC} method has two dependences that must be set before the component separation is performed: the set of needlets and the internal linear combination (ILC) bias $b$. These parameters control the localization that is made by {\tt GNILC} when calculating the covariance matrices. Needlets determine the location in harmonic and real (or pixel) space. The most appropriate set of needlets for BINGO maps is the one that combines the strengths from a mild localization at low multipoles and a fine localization at high multipoles. 
Figure \ref{fig:needlet} shows the set adopted in our analysis.
The ILC bias, however, is not a totally free parameter, as its increase leads to an increase in the artificial anticorrelation between the component of interest and the contaminants. It should, therefore, be made as small as possible without increasing the resulting localized area of the sky and computational processing too much.

In order to test the sensitivity of the method as a function of the simulation parameters, we attempted to recover the power spectra for scenarios with a different number of feeds (double-rectangular 28 and 56 and hexagonal 31), different $1/f$ noise correlations and different observation times.
%we performed power spectra recovery attempts for scenarios with different focal plane arrangements and numbers of feeds (Double-rectangular 28, Hexagonal 31 and Double-rectangular 56).
To have a quantitative measure of the ability of the {\tt GNILC} method, we compare the results with the standard PCA method, which is commonly used in \textsc{Hi} IM simulations \citep{Sazy15, Alonso15}. The PCA method consists of the transformation of the independent maps of each frequency channel into orthogonal modes of the covariance matrix between frequencies. This method relates the foreground components to the eigenvectors with the largest variance modes, which are called the principal components. These principal components are then removed from the data.

As a way to quantify the recovery performance of the methods, the average absolute difference between the input $C_{\ell}^{s}$ and the reconstructed power spectrum  $C_{\ell}^{R}$  normalized by the input power spectrum is considered:

\begin{equation}
{N_{Rec}} = \frac{1}{{{N_\ell }{n_{ch}}}}\sum\limits_i^{{n_{ch}}} {\sum\limits_\ell ^{{\ell _{\rm{max}}}} {\left| {\frac{{C_\ell ^R\left( {{\nu _i}} \right) - C_\ell ^S\left( {{\nu _i}} \right)}}{{C_\ell ^S\left( {{\nu _i}} \right)}}} \right|} }\,, 
\label{eq:Ngnilc}
\end{equation}

\noindent where $N_{\ell}$ is the total number of multipoles considered and $\nu_{i}$ specifies a frequency channel. The ideal scenario would be when $N_{Rec}$ = 0, indicating a perfect \textsc{Hi} plus noise map reconstruction.
We summarize, in Table \ref{tab:n_results}, the values of $N_{Rec}$ obtained for each configuration. All the results shown have been obtained with the same value of the ILC bias (0.01). For the choice of needlets, we used a set that combines, as mentioned before, the strengths from a mild localization at low multipoles and a fine localization at high multipoles. Regarding the standard PCA, the recovery of the \textsc{Hi} signal with the smallest bias is obtained when we choose the number of principal components to be equal to 3.

%Note that for these simulations we have set the $1/f$ spectral index $\beta$ = 0.25 according to the investigation of $1/f$ noise in \textsc{Hi} IM by \cite{Harper18b}.
Measurements made with a digital back-end for an earlier version of the BINGO receiver yielded $\beta \approx 0.25$, which we use as our fiducial value.
The power spectral density of the spectral gain fluctuations is modeled as a power law and characterized by the parameter $\beta$. Small values of $\beta$ ($< 0.25$), or high correlation, are preferred, as this makes it easier to remove the 1/f fluctuations using current component separation techniques. The value for $\beta$ is heavily dependent on the receiver setup. 
For our analysis, we considered the frequency that is closest to the middle point of the BINGO band.

\begin{figure}
    \centering
        \includegraphics[width=\hsize]{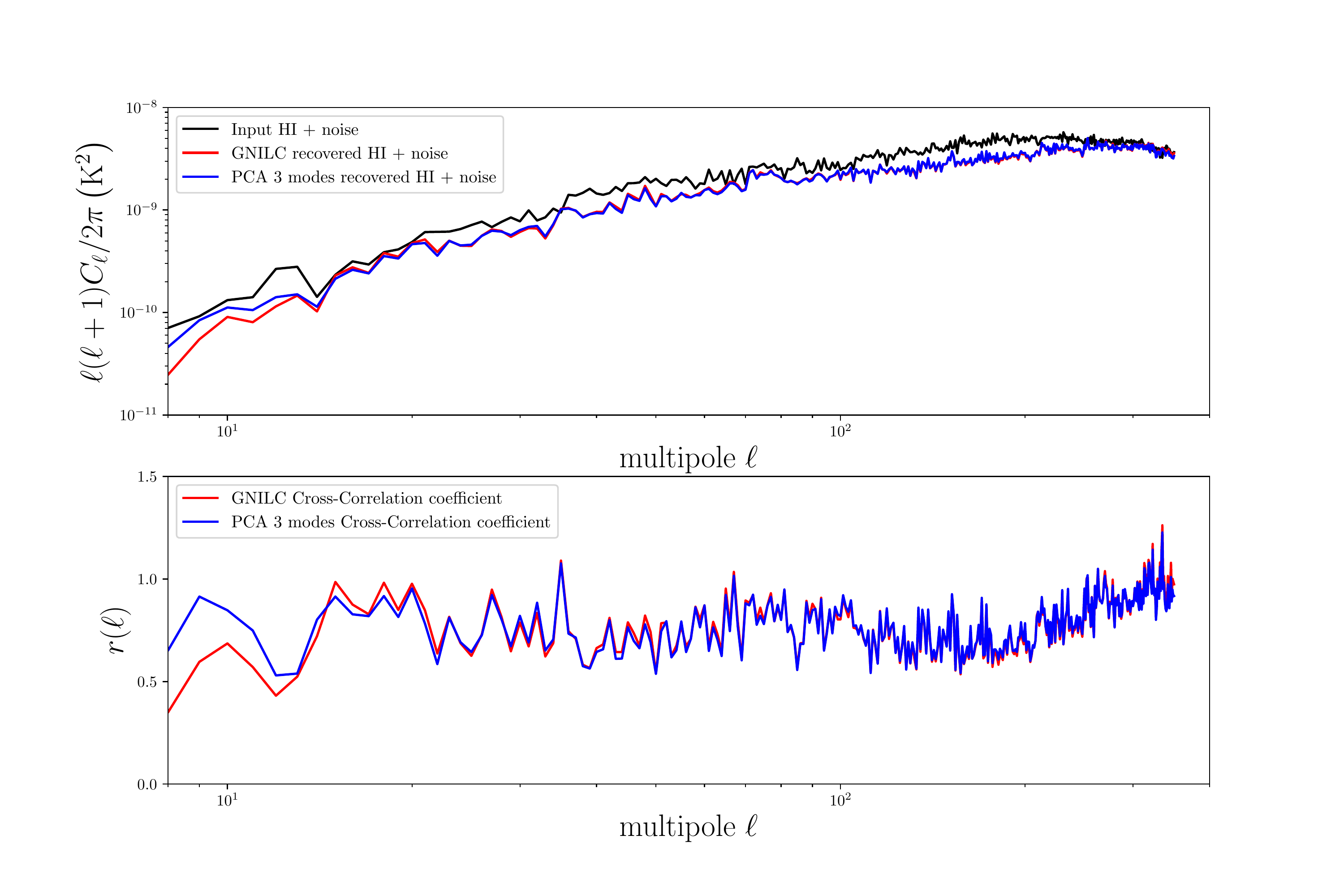}
    \caption{Component separation results using the simulation pipeline. $Top$: Angular power spectra for the input \textsc{Hi} plus noise signal ($black$), the {\tt GNILC} recovered \textsc{Hi} plus noise signal ($red$), and the PCA recovered \textsc{Hi} plus noise signal with three modes removed ($blue$) at $\approx$ 1.1\,GHz. For this particular channel and configuration (double-rectangular, 28 feeds, one-year observation time, and $\beta$ = 0.25), $N_{Rec}$ ({\tt GNILC}) equals 21.43$\%$. $Bottom$: Cross-correlation coefficient $r(\ell)$ among the recovered signals ({\tt GNILC} and PCA) and the input signal.}
    \label{fig:gnilc1}
\end{figure}

\begin{table}
\centering
\caption{Average normalized absolute difference between the input power spectrum and the recovered power spectrum of the \textsc{Hi} plus noise signal ($N_{Rec}$) for the central frequency ($\approx$ 1.1\,GHz) with $\beta$ = 0.25. ``D.R.'' stands for ``double rectangular'' and ``HEX'' for `hexagonal arrays.''}
%\vspace{2pc}
\resizebox{9cm}{!}{
\begin{tabular}{c||c||c||c} \toprule
{\bf Number of feeds}       & 28 (D.R.)    & 31 (HEX)       & 56 (D.R.)        \\  \midrule       
{\bf Integration time (y)}  &  1 \space \space \space \space \space  \space  \space \space \space 2 \space \space \space \space \space  \space  \space \space \space 5   & 1         & 1  \\ \midrule   
{\bf $N_{Rec}$ (\%)} {\tt GNILC}   & 21.43 \space  \space  \space \space \space 20.12 \space \space \space  \space  \space   14.31 &  21.05    & 19.6  \\ \midrule
{\bf $N_{Rec}$ (\%)} PCA   & 20.37 \space  \space  \space \space \space 18.93 \space \space \space  \space  \space   12.87   &  20.13    & 18.2  \\ \bottomrule
    \bottomrule
\end{tabular}
}
\label{tab:n_results}
\end{table}

Figure \ref{fig:gnilc1} shows the power spectra of the input \textsc{Hi} plus noise signal and the {\tt GNILC} and PCA recovered \textsc{Hi} plus noise signal, relative to the scenario with 28 feeds and one-year observation time. In the bottom panel of Fig. \ref{fig:gnilc1} we plot the cross-correlation coefficient, defined as

\begin{equation}
r\left( \ell  \right) = \frac{{C_\ell ^R\left( {{\nu _i}} \right)C_\ell ^S\left( {{\nu _i}} \right)}}{{C_\ell ^S\left( {{\nu _i}} \right)C_\ell ^S\left( {{\nu _i}} \right)}}.
\end{equation}

\begin{figure*}[ht]
\begin{center}
        \includegraphics[width=.75\textwidth]{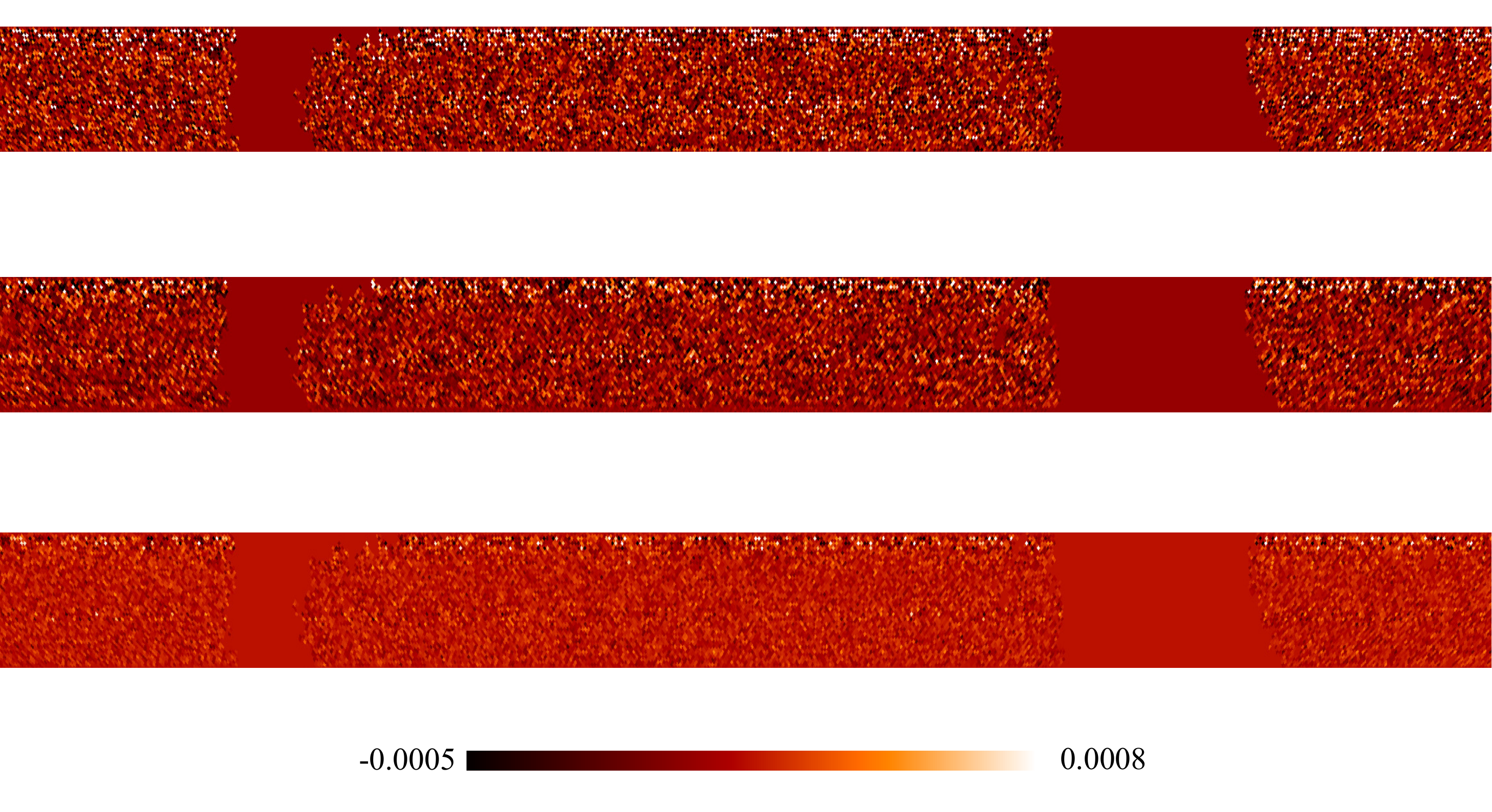}
        \end{center}
    \caption{BINGO \textsc{Hi} plus noise maps with a Galactic mask applied at $\approx$ 1.1\,GHz (double-rectangular, 28 feeds, one-year integration time, and $\beta$ = 0.25). In the $top$ panel we have the input map, in the $middle$ panel the {\tt GNILC} recovered map, and in the $bottom$ panel the residuals map (temperatures are given in K).}
    \label{fig:stripes}
\end{figure*}

The cross-correlation coefficient is a complementary way to measure the signal recovery sensitive to scale-dependent signal loss.
Depending on the number of principal components that are removed, the PCA either underestimates the \textsc{Hi} power spectrum or is contaminated by residual foregrounds. The best result was obtained with 3 modes removed.
We can see that the recovery of the two methods is comparable throughout the entire range of multipoles $\approx$ 12 $<$ $\ell$ $<$ 330. The performance also depends on the range of angular scales. The cross-correlation coefficient is $<$ 1 throughout almost the entire range of multipoles, while on smaller scales its value increases, showing that the recovered spectrum is contaminated. The {\tt GNILC} recovered and residuals maps covering the BINGO sky region for this scenario are shown in Fig. \ref{fig:stripes}.

\begin{figure}
    \centering
        \includegraphics[width=\hsize]{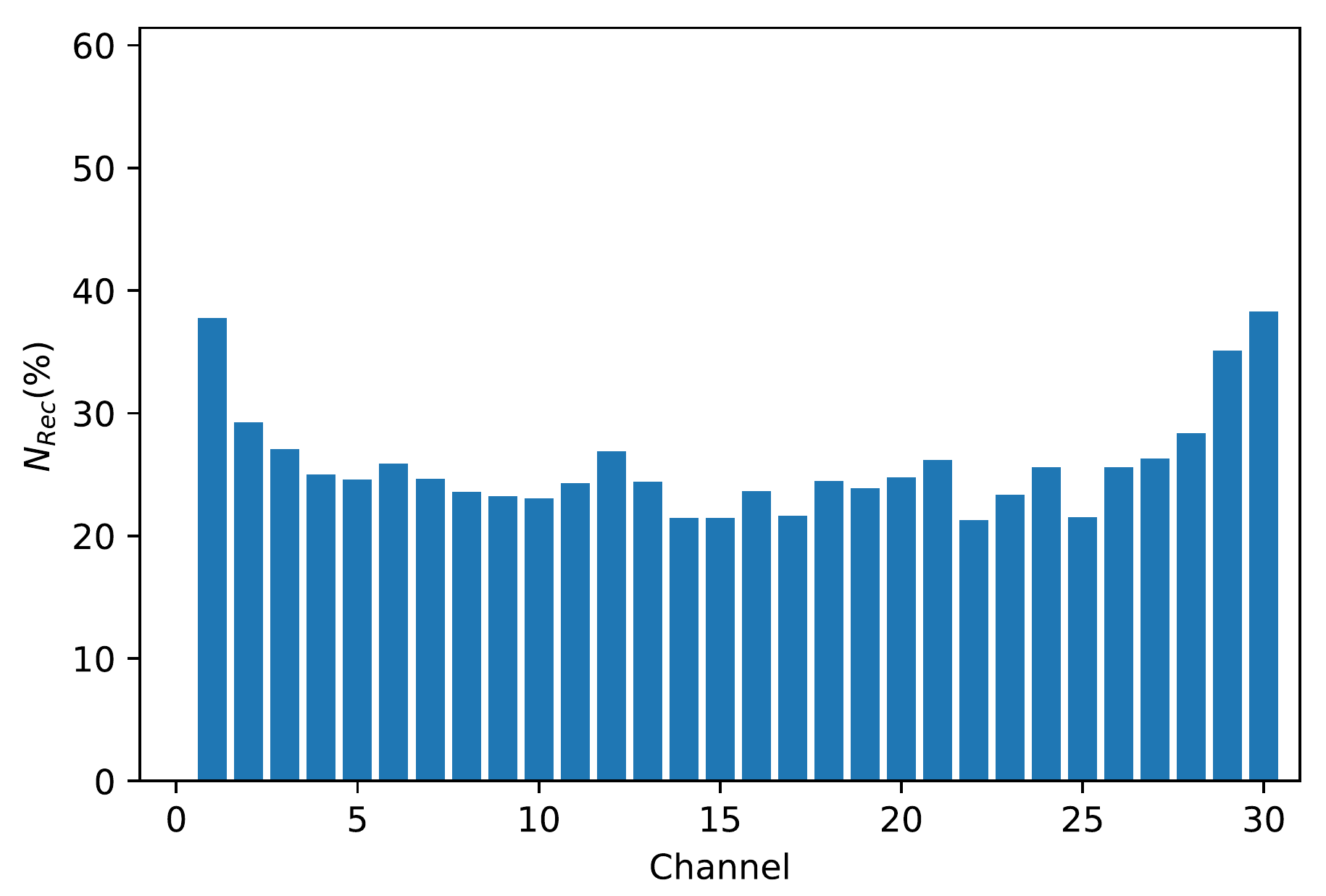}
    \caption{Histogram showing the average absolute difference between the input and the recovered \textsc{Hi} power spectrum normalized by the input \textsc{Hi} power spectrum ($N_{Rec}$) obtained with {\tt GNILC} for each channel of the BINGO frequency band.}
    \label{fig:hist}
\end{figure}

For the three different configurations, the input power spectrum is recovered with good accuracy, with an improvement as the number of feeds is increased (Table \ref{tab:n_results}). The results show that the component separation method does not affect significantly the wanted signal statistics, but it underestimates the power spectrum (Fig. \ref{fig:gnilc1}). We show in Fig. \ref{fig:hist} the values of $N_{Rec}$ ({\tt GNILC}) obtained for each channel. Performances are worse at the edges of the frequency band, where there is less freedom for the method to fit for the independent components of emission without compromising the reconstruction of the wanted signal.

We finally summarize the {\tt GNILC} performance for different multipole ranges in Table \ref{tab:range} (scenario with 28 feeds and one-year observation time). The difference between the total signal and the \textsc{Hi} plus noise signal is largest in the middle part of the interval, where the Galactic foregrounds are more intense. For smaller scales ($\ell$ $>$ 200) instead, we can recover the cosmological \textsc{Hi} plus noise power spectrum with very good accuracy. This happens because in this range of scales the difference in power between the total signal and the wanted signal is smaller. It is worth noting that the {\tt GNILC} performance may be adjusted by changing the ILC bias, whose value represent the localization in the real space. There is a direct dependence on it when the localization in the spherical space is not fine enough. We are investigating this aspect for a future publication.

\begin{table}
        \centering
        \caption{$N_{Rec}$ values for different ranges of multipoles (double-rectangular, 28 feeds, one-year observation time, and $\beta$ = 0.25) obtained with {\tt GNILC} at $\approx$ 1.1 GHz.}
        \label{tab:range}
        \begin{tabular}{cc} \toprule % four columns, alignment for each
                \hline
                {\bf Range of multipoles} & $N_{Rec}$ \\ 
                \hline
                \toprule
                15--30 & 0.18 \\ \midrule
                30--60 & 0.28 \\ \midrule
                60--90 & 0.19 \\ \midrule
                90--120 & 0.23 \\ \midrule
                120--150 & 0.30 \\ \midrule
                150--180 & 0.31 \\ \midrule
                180--210 & 0.31 \\ \midrule
                210--240 & 0.22 \\ \midrule
                240--270 & 0.14 \\ \midrule
                270--300 & 0.12 \\ \midrule
                300--330 & 0.08 \\ \bottomrule
                \bottomrule
                \hline
        \end{tabular}
\end{table}

\begin{figure}
    \centering
        \includegraphics[width=\hsize]{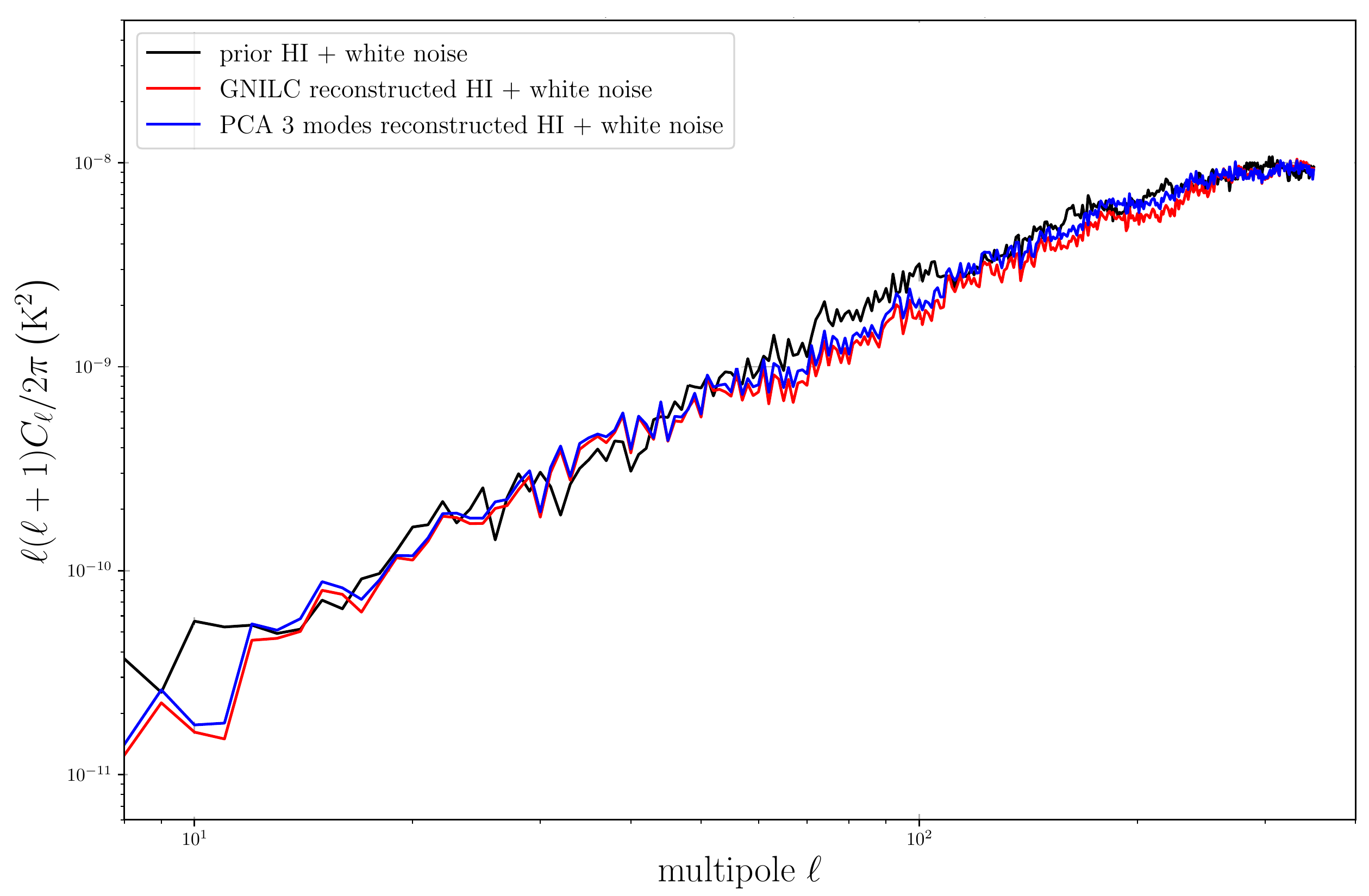}
    \caption{Angular power spectra for the input \textsc{Hi} plus noise signal ($black$), the {\tt GNILC} recovered \textsc{Hi} plus noise signal ($red$), and the PCA recovered \textsc{Hi} plus noise signal with three modes removed ($blue$) at $\approx$ 1.1\,GHz. For this particular channel and configuration (double-rectangular, 28 feeds, one-year observation time, and $\beta$ = 0.001), $N_{Rec}$ ({\tt GNILC}) equals $\approx$ 12 $\%$. The recovery improvement, due to the increased $1/f$ noise correlation, is clear.}
    \label{fig:gnilc1_corr}
\end{figure}

In this context, we used the {\tt GNILC} method as well to investigate the impact of $1/f$ noise correlation in the reconstruction of \textsc{Hi} plus thermal noise signal. Table \ref{tab:1/f} shows the results when varying $\beta$. It can be noted that the degree of $1/f$ noise correlation between the receivers channels affects the recovery of the input power spectrum. According to these results, the {\tt GNILC} method is capable of recovering the spectrum but with different performances.  
Our analysis shows that, with real data, the $1/f$ noise contribution will be the more challenging contaminant to be removed. Figure \ref{fig:gnilc1_corr} shows the {\tt GNILC} and PCA recovered \textsc{Hi} plus noise power spectra when there is a high degree of $1/f$ noise correlation ($\beta$ = 0.001, D.R. 28 feeds, one-year observation time). As expected, an improvement in the {\tt GNILC} performance with an increase of correlation in frequency is found. The $1/f$ noise can be sufficiently reduced to be significantly lower than the thermal noise based on a one-year observation time with the same instrumental parameters described in Table \ref{parameters_simul}. Further improvements could also be achieved by using more advanced map-making codes. There are some discussed in the literature (e.g., \citealp{Natoli01,de_Gasperis_2016}) that should be able to suppress in part the $1/f$ noise. 
For the real observations the 1/f noise effect in the data can be diminished in two ways: either by using very stable receivers so that the knee frequency is reduced or by increasing the scanning speed so that the signal is shifted to higher frequencies. For a static telescope like BINGO this implies that the receiver gains need to be as stable as possible.
%$1/f$ noise should be expected to be more challenging to remove than implied by these simulations.

\begin{table}
        \centering
        \caption{$N_{Rec}$ for different $\beta$ values (double-rectangular, 28 feeds, one-year observation time) obtained with {\tt GNILC} at $\approx$ 1.1\,GHz.}
        \label{tab:1/f}
        \begin{tabular}{ccr} \toprule% four columns, alignment for each
                \hline
                {\bf Spectral index} & $N_{Rec}$\\
                \hline
                \toprule
                0.001 & 0.12\\ \midrule
                0.12 & 0.2 \\ \midrule
                0.25 & 0.21 \\ \midrule
                0.6    & 0.32 \\ \bottomrule
                \bottomrule
                
                \hline
        \end{tabular}
\end{table}

%%%%%%%%%%%%%%%%% NEW .ION %%%%%%%%%%%%%%%%%%

\section{Conclusions}

BINGO is a transit telescope designed to make the first detection of BAOs at radio wavelengths, through the detection of the 21 cm \textsc{Hi} emission and, consequently, the construction of a 3D map of total matter distribution. In this way, the telescope will provide independent cosmological data to complement optical surveys at the redshift interval $0.127 < $ z $ < 0.449$. 

% Thanks to a good location for astronomical observations at these frequencies, the wide field-of-view, the configuration and optical design with multiple beams, the very large focal plane and the way of continuous and repeated observation over the same region of the sky, BINGO will be also a powerful instrument capable of detecting transient radio phenomena, including fast radio bursts ($\sim$ 1 every 4 days), by just adjusting integration times and frequency channels widths. 

% The telescope is based on simplicity and low cost: the mirrors are static and the receiver system is standard (no cryogenics and off-the-shelf components). The construction operations of the telescope are scheduled for the beginning of 2022.

We have performed an optimization analysis to determine the best solution for the BINGO focal plane arrangement. Two different arrays have been considered, and other possible configurations can be found in previous works \citep{Battye13, Sazy15}. According to the results obtained in this work, the optimal solution for the BINGO focal plane design is the double-rectangular array. This design concept will use a large structure to mount an array of 28 horns, allowing the displacement of each horn along the vertical axis (elevation axis). The structure will be able to host, in the future, up to 56 horns.

In this work we have presented, for the first time, results from the application of the {\tt GNILC} method to a set of end-to-end simulations generated with the IM pipeline developed by the BINGO collaboration.
%BINGO collaboration IM pipeline. 
The foreground cleaning method did show satisfactory results with the parameters adopted in this work. 
The simulations were carried out with $n_{ch}$ = 30 due to computational processing time constraints. This parameter has a direct influence on the efficiency of {\tt GNILC}.  With more frequency channels, there is more freedom for the method to fit for the independent components of emission (the foregrounds) without compromising the reconstruction of the wanted signal. The telescope is expected to work with more channels during the data collection, and this will improve the capability of the foreground cleaning method. 

In line with the results of this work, we can say that {\tt GNILC} can reconstruct the \textsc{Hi} plus noise signal, for the IM BINGO experiment, with an (absolute) accuracy of $\approx $12\% after five years of observation. This result may be improved by adding more horns to the telescope. The reconstruction is good even in the presence of systematics, such as $1/f$ noise, which was not considered in previous works \citep{Olivari16}. This has been an encouraging result since this systematic will be present, to a great extent, in the real data and is the most complex contaminant to remove. Map-making methods can help to reduce the contribution of $1/f$ noise but at present are not optimized for our \textsc{Hi} IM simulations. This is an important point because it indicates that it is preferable, where possible, to try to suppress $1/f$ noise in hardware, such as with the BINGO correlated channels, than to rely entirely on post-processing in software.

 We note that the current implementation of {\tt GNILC} in our analysis relies on a simple choice of the needlet bands, without any attempt at optimizing localization inside the small sky area observed by BINGO. The set was determined with equal intervals between peaks, with prioritization for high multipoles ($\ell$ $>$ 100), and is not definitive. In a future paper, for a more accurate cosmological signal recovery, a study to optimize their quantity and their distribution will be required. We intend to optimize the needlet localization inside the sky area observed by BINGO and anticipate some improvement of {\tt GNILC} over a simple PCA since such localization will allow the number of principal components estimated by {\tt GNILC} to vary inside the BINGO area. This is expected because {\tt GNILC} estimates the number of principal components locally across the sky and, thus, allows the effective number of principal components to vary across the sky, while a PCA assumes a fixed number of principal components (e.g., three), all over the sky, which is a rough approximation given the expected variation in non-Gaussian foregrounds and signal-to-noise across the sky area observed by BINGO.

We can summarized the work presented in this paper as follows:
\begin{itemize}
    \item Simulated cosmological signal with {\tt FLASK};\\ 

    \item Foreground modeling with galactic synchrotron, free-free, and AME;\\

    \item Simulated instrumental noise  (thermal noise plus $1/f$ noise);\\
    
    \item Sky masking for different observational strategies;\\
    
    \item Cosmological \textsc{Hi} signal recovery with {\tt GNILC};\\
    
    \item Comparison of simulated and recovered angular power spectra.

\end{itemize}

%As the telescope construction proceeds, we will keep improving the work to obtain a more efficient separation of components, and intend to address the following items in the near future:

As the telescope construction proceeds, we will also work on the improvement of this pipeline to include other component separation methods and intend to address the following items in the near future: 
\begin{itemize}

\item A more careful estimation of the errors in the recovered maps. One way to accomplish this is through Monte Carlo runs of different realizations of the observed sky. The main problem here is that it will require the simulation of a large number of different foreground skies, much larger than the number of available models that exist in the literature. A solution to this problem could be the use of some toy models for the spatial and frequency distribution of the different types of foreground emissions.\\

  \item Testing of different sets of needlets;\\

  \item Refining the $1/f$ noise model;\\

  \item Including more realistic terrestrial RFI  measurements in the simulations;\\

  \item Including extragalactic point radio sources maps in the simulations.

\end{itemize}

The accomplishment of these points will allow us to produce more realistic simulations, which are a valuable tool for testing and verification of the quality of the component separation steps during the BINGO data analysis.

\section*{Acknowledgements}
The BINGO project is supported by FAPESP grant 2014/07885-0. 
V. L. acknowledges the postdoctoral FAPESP grant 2018/02026-0. 
C.A.W. acknowledges CNPq through grant 313597/2014-6. 
T.V. acknowledges  CNPq  grant  308876/2014-8. 
F.B.A. acknowledges the UKRI-FAPESP grant 2019/05687-0, and FAPESP and USP for Visiting Professor Fellowships where this work has been developed.
E. M. acknowledges a Ph.D. CAPES fellowship. 
F. V. acknowledges a CAPES M.Sc. fellowship. 
C.P.N. thanks S{\~a}o Paulo Research Foundation (FAPESP) for financial support through grant 2019/06040-0,
K.S.F.F. thanks S{\~a}o Paulo Research Foundation (FAPESP) for financial support through grant 2017/21570-0.
Support from CNPq is gratefully acknowledged (E.A.). 
R.G.L. thanks  CAPES (process 88881.162206/2017-01) and the Alexander von Humboldt Foundation for the financial support. 
A.R.Q., F.A.B., L.B., and M.V.S. acknowledge PRONEX/CNPq/FAPESQ-PB (Grant no. 165/2018). 
M.P. acknowledges funding from a FAPESP Young Investigator fellowship, grant 2015/19936-1. 
J.Z was supported by IBS under the project code, IBS-R018-D1. 
L.S. is supported by the National Key R\&D Program of China (2020YFC2201600).
M.R. acknowledges funding from the European Research Council Grant CMBSPEC (No. 725456).
A.A.C. acknowledges financial support from the China Postdoctoral Science Foundation, grant number 2020M671611. 
 B. W. and A.A.C. were also supported by the key project of NNSFC under grant 11835009.
Some of the results in this paper have been derived using the HEALPix package (http://healpix.sourceforge.net) \citep{Gorski05}.
This research made use of Astropy,\footnote{http://www.astropy.org} a community-developed core Python package for Astronomy \citep{astropy:2018}, NumPy \citep{harris2020array} and Healpy \citep{Zonca2019}.

%%%%%%%%%%%%%%%%%%%%%%%%%%%%%%%%%%%%%%%%%%%%%%%%%%

%%%%%%%%%%%%%%%%%%%% REFERENCES %%%%%%%%%%%%%%%%%%

% The best way to enter references is to use BibTeX:

\bibliographystyle{aa}
\bibliography{aa_cit.bib} % if your bibtex file is called example.bib

% Alternatively you could enter them by hand, like this:
% This method is tedious and prone to error if you have lots of references
%\begin{thebibliography}{99}
%\bibitem[\protect\citepauthoryear{Author}{2012}]{Author2012}
%Author A.~N., 2013, Journal of Improbable Astronomy, 1, 1
%\bibitem[\protect\citepauthoryear{Others}{2013}]{Others2013}
%Others S., 2012, Journal of Interesting Stuff, 17, 198
%\end{thebibliography}

%%%%%%%%%%%%%%%%%%%%%%%%%%%%%%%%%%%%%%%%%%%%%%%%%%

% Don't change these lines
%\bsp   % typesetting comment
\label{lastpage}
\end{document}